\documentclass[journal]{IEEEtran}

\usepackage[pdftex]{graphicx}
\graphicspath{{../pdf/}{../jpeg/}}
\DeclareGraphicsExtensions{.pdf,.jpeg,.png}
\usepackage[cmex10]{amsmath}
\usepackage{amsfonts}
\usepackage{amsthm}
\usepackage[english]{babel}
\usepackage{graphicx,color}
\usepackage{cite}
\usepackage{enumerate}
\usepackage{blindtext}
\usepackage{tcolorbox}
\usepackage{xcolor}
\usepackage{lipsum}
\usepackage{algorithm}
\usepackage{algorithmic}
\usepackage{array} 
\usepackage{tabularx}
\usepackage{makecell}
\usepackage{subcaption}
\newtheorem{proposition}{Proposition}
\usepackage[T1]{fontenc} 

\newtheorem{theorem}{Theorem}
\usepackage{amssymb}

\begin{document}
	\title{\huge{Max-Min Fair Energy-Efficient Beamforming Design for Intelligent Reflecting Surface-Aided SWIPT Systems with Non-linear Energy Harvesting Model}}
	
	\author{Shayan Zargari,~Ata Khalili,~\textit{Member,~IEEE},~Qingqing Wu,~\textit{Member,~IEEE},~Mohammad Robat Mili, and  Derrick Wing Kwan Ng,~\textit{Fellow,~IEEE}
		\thanks{Shayan. Zargari is with the School of Electrical Engineering,~Iran University of Science and Technology,~Tehran,
			Iran (e-mail: shayanzargari66@ gmail.com). Ata Khalili is with Electronics Research Institute,~Sharif University of Technology,
			Tehran,~Iran. Ata Khalili is also with the Department of Electrical and Computer Engineering Tarbiat Modares University,~Tehran,
			Iran (e-mail: ata.khalili@ieee.org). Qingqing Wu is with the State Key Laboratory of Internet of Things for Smart City and Department of Electrical and Computer Engineering,~University of Macau,~Macao 999078,~China (e-mail: qingqingwu@um.edu.mo). D. W. K. Ng is with the School of
			Electrical Engineering and Telecommunications,~University of New South
			Wales,~Sydney,~NSW 2052,~Australia (e-mail: w.k.ng@unsw.edu.au).}}    
	\maketitle
	
	\begin{abstract}
		This paper considers an intelligent reflecting surface (IRS)-aided simultaneous wireless information and power transfer (SWIPT) network,~where multiple users decode data and harvest energy from the transmitted signal of a transmitter.~The proposed design framework exploits the cost-effective IRS to establish favorable communication environment to improve the fair energy efficient.~In particular, we study the max-min energy efficiency (EE) of the system by jointly designing the transmit information and energy beamforming at the base station (BS),~phase shifts at the IRS,~as well as the power splitting (PS) ratio at all users subject to the minimum rate, minimum harvested energy, and transmit power constraints. The formulated problem is non-convex and thus challenging to be solved. We propose two algorithms namely penalty-based and inner approximation (IA)-based to handle the non-convexity of the optimization problem.~As such,~we divide the original problem into two sub-problems and apply the alternating optimization (AO) algorithm for both proposed algorithms to handle it iteratively.~In particular, in the penalty-based algorithm for the first sub-problem, the \textcolor{blue}{semi-definite} relaxation (SDR) technique, difference of convex functions (DC) programming, majorization-minimization (MM) approach, and fractional programming theory are exploited to transform the non-convex optimization problem into a convex form that can be addressed efficiently.~For the second sub-problem,~a penalty-based approach is proposed to handle the optimization on the phase shifts introduced by the IRS with the proposed algorithms.~For the IA-based method, we optimize jointly beamforming vectors and phase shifts while the PS ratio is solved optimally in the first sub-problem. Simulation results verify the effectiveness of the IRS, which can significantly improve the system EE as compared to conventional benchmarks schemes and also unveil a trade-off between convergence and performance gain for the two proposed algorithms.
	\end{abstract}
	
	\section{Introduction}
	The fifth-generation (5G) communications attain numerous gains in both spectral efficiency (SE) and energy efficiency (EE) by exploiting different advanced technologies \cite{1}. Nevertheless,~these technologies generally require high power dissipation and high implementation cost,~which have constituted the bottleneck in designing practical systems \cite{2}. For instance,~the Internet-of-Things (IoT) devices are limited by the battery capacity,~which has become a bottleneck of such networks.~Thus,~a scalable and sustainable solution is needed to achieve ubiquitous connectivity and constant energy supply for these devices in 5G wireless networks and beyond \cite{3}.~Recently,~intelligent reflecting surface (IRS) has been firstly proposed as a promising technology to achieved high beamforming gain with significantly reduced energy consumption and hardware cost \cite{77}, \cite{7}--[10].~These surfaces can be installed on the outsides of buildings to provide energy-efficient wireless communications by reducing the transmission power of the base station (BS). Especially,~an IRS consists of a large number of adjustable passive units mounted on a planar array,~which are coordinated smartly to reflect signals in order to establish more favorable wireless propagation channels and achieve more reliable communication.~These passive elements are indeed cost effective and energy efficient,~which can intelligently adjust the phases to steer the incident signals independently \cite{77, Wu44}.~Consequently,~the reflected signal can be combined coherently at the desired receiver to boost the signal-to-interference-plus-noise (SINR) \cite{16}.~Therefore, IRS is a disruptive technology for making our current “dumb” environment intelligent, which can potentially benefit a wide range of vertical industries in 5G/6G such as transportation, manufacturing, smart city, etc \cite{Tutorial}.~In particular,~we would like to investigate if the deployment of the IRS can be beneficial in a power splitting (PS)-based simultaneous wireless information and power transfer (SWIPT) system and how the performance of such a system is compared to a conventional one that does not exploit IRS.
	\subsection{Related Works}
	The related works on this topic can be classified into three groups,~in the following,~we discuss the topics individually,~namely,~i) IRS systems; ii) SWIPT systems; iii) IRS-aided SWIPT systems.
	\begin{table*}[ht]
		\caption{Summary of Related Works}
		\label{related work}
		\centering
		\begin{tabular}{|c|c|c|c|c|c|}\hline
			{\bf Ref.} &  \thead{\bf Type of system} & \thead{\bf Objective Function} & {\bf Constraints} & \thead{\bf Solution Approaches} \\
			\hline 
						\cite{16} & \thead{MISO-IRS} & \thead{Minimizing total
				transmit power} & \thead{Individual SINR constraints,\\phase shifts at the IRS} & \thead{SDR,\\ AO algorithm} \\
			\hline
			\cite{8} & \thead{MISO-IRS} & \thead{Maximizing EE} & \thead{Individual link budget of receivers,~\\ phase shifts at the IRS}& \thead{Gradient descent search,~\\Sequential fractional programming} \\
			\hline
			\cite{9} & \thead{MISO-IRS with \\ ZF transmission} & \thead{Maximizing sum-rate} & \thead{Transmission power at the BS,~\\ QoS constraints for receivers,~\\ phase shifts at the IRS} & \thead{Majorization-minimization\\
				method} \\
			\hline
			\cite{13} & \thead{Multigroup
				multicast\\ MISO-IRS} & \thead{Maximizing \textcolor{blue}{sum-rate} } & \thead{Precoding matrix at the BS,~\\ phase shifts at the IRS} & \thead{Majorization-minimization\\
				method}\\
				\hline
		\cite{Schoberr,Green_IRS}& 	\thead {MISO-IRS} & 	\thead{Minimizing total
			transmit power} & 	\thead{QoS constraints for receivers,\\ phase shifts at the IRS} &	 \thead{
			SDR,\\
			DC programming,\\
			AO algorithm,
			\\penalty-based method, \\
			IA-based method}\\
			\hline
			\cite{20} &  \thead{ MISO-IRS-SWIPT\\ with separated receivers\\ and linear EH model}& \thead{Maximizing weighted \textcolor{blue}{sum-power}} & \thead{Individual SINR constraints,\\phase shifts at the IRS} & \thead{SDR,\\ AO algorithm} \\
			\hline
			{\cite{22}} & \thead{ MISO-IRS-SWIPT\\ with separated receivers\\ and linear EH model}& \thead{Minimizing total
				transmit power} & \thead{{Individual SINR constraints,} \\ {EH constraints},\\phase shifts at the IRS} & \thead{Penalty-based method,\\Block coordinate descent
				algorithm} \\
			\hline
			{\cite{23}} &\thead{ MISO-IRS-SWIPT\\ with separated receivers\\ and linear EH model} & \thead{Max-min harvested power} & \thead{Transmission power at the BS,~\\ individual SINR constraints,~\\ phase shifts at the IRS} & \thead{SDR,\\AO algorithm} \\
			\hline
			\cite{24} & \thead{ MIMO-IRS-SWIPT\\ with separated receivers\\ and linear EH model} & \thead{Maximizing weighted sum-rate} & \thead{Transmission power at the BS,\\EH constraints,~\\ phase shifts at the IRS}& \thead{Block coordinate descent algorithm,\\ AO algorithm}\\
			\hline
		{\cite{Zargari}} &	{ \thead{ MISO-IRS-SWIPT\\ with co-located receivers\\ and linear EH model} }& 	{\thead{Maximizing EE indicator}} &	{ \thead{Transmission power at the BS,\\ phase shifts at the IRS}}&		{\thead{Fractional and sequential programming,\\majorization-minimization,\\SDR,\\manifold method,\\AO algorithm}}\\
			\hline
			\thead{Our work} & 	\thead{ MISO-IRS-SWIPT\\ with co-located receivers\\ and non-linear EH model} & 	\thead{Max-min individual EE} & 	\thead{Transmission power at the BS,~\\EH constraints,\\individual SINR constraints,\\ PS ratio at each user,~\\phase shifts at the IRS} & 	\thead{Fractional and sequential programming,\\
				SDR,\\majorization-minimization,\\
				DC programming,\\
				AO algorithm,
				\\penalty-based method, \\
			IA-based method}\\
			\hline
		\end{tabular}
	\end{table*} 
	
	\textit{i) IRS systems:}
	To fully exploit the performance gain promised by the IRS,~a plethora of research studies has widely investigated IRS-assisted wireless communication \cite{16}--\cite{Green_IRS}.~The authors in \cite{16} investigated minimizing the total transmitted power at the BS in an IRS-assisted multiple-input single-output (MISO) network.~Specifically,~for the single-user case,~the fundamental squared power gain has been unveiled.~For the multi-user case,~it has been demonstrated that SINR performance can be considerably enhanced by jointly optimizing the beamforming vectors at the BS and the phase shifts at the IRS.~Besides,~a downlink MISO multi-user scenario with the aid of the IRS was investigated in \cite{8},~where the transmit power and passive phase shifts at the IRS were jointly designed to provide an energy-efficient communication system.~The authors in \cite{9} considered the IRS in a wireless communication system to improve the sum-rate under a multi-user scenario.~Moreover,~a suboptimal solution based on the zero-forcing (ZF) precoding was considered at the BS to optimize the passive elements at the IRS.~Also,~the authors in \cite{11} obtained an upper bound for the ergodic SE,~which highly depends on the passive elements at the IRS.~In particular,~the ergodic SE was maximized by designing the phase shifts under statistical channel state information (CSI).~Furthermore,~\cite{13} proposed an IRS-assisted multigroup multicast network in which the sum data rate of all groups was maximized by aligning the passive elements at the IRS.~Minimizing the transmit power with joint optimization of transmit precoding and phase shifts was investigated in \cite{Wu11} under single-user and multiple user scenarios.~In particular,~it was shown that discrete phase shifts achieved by quantizing the optimized continuous phase shifts almost obtain optimal performance in the single-user scenario.~For the multi-user scenario,~performance degradation is not significant since the co-channel interference is serious.~The application of the IRS in cognitive radio communication systems was investigated in \cite{Wu22} in order to share the spectrum between a secondary user (SU) link and a primary user link.~In particular,~the SU's rate was maximized by jointly designing the phase shifts at the IRS and the SU transmit power.~In \cite{Schoberr} and \cite{Green_IRS}, a new algorithm based on the inner approximation (IA) was proposed to minimize the total transmission power in an IRS-aided MISO system at the cost of lower convergence speed.~However, they showed that the algorithm can converge to a stationary point,~unlike existing algorithms.

	\textit{ii) SWIPT \textcolor{blue}{systems}:} Energy consumption is an important issue that needs to be addressed in future wireless systems.~In the meantime,~SWIPT technology was proposed to improve EE as well as energy fairness \cite{18}--[23].~More precisely,~this technology allows information decoding (ID) and energy harvesting (EH) simultaneously for each user based on the received signals.~To realize such a concept in practice by taking into account the sensitivities of various receivers,~there exists two practical receiver architectures in SWIPT-based networks,~named as PS and time switching (TS)\cite{18}. For the PS architecture,~a single user can employ a splitter to divide the received signal into two distinct power streams with a specified ratio such that one stream is for the ID,~and the other one is for EH\cite{19-1}. The authors in \cite{21} investigated a multi-user MISO network for two kinds of receivers,~i.e.,~information decoder receivers (IDRs) and energy harvester receivers (EHRs).~Especially,~they aimed at maximizing the weighted sum of total harvested power.~\cite{222} studied co-located receivers where a single user employs a PS architecture.~In particular,~both optimal and suboptimal algorithms were designed to minimize the total transmission power at the BS.~Note that in SWIPT systems,~path loss in long-range communication has a severe effect on the amount of harvested energy at the user side,~creating a performance bottleneck.~Hence,~IRS and the multiple antennas can be exploited jointly in the PS-based SWIPT-aided network to enhance the transmission power efficiency by exploiting the significant beamforming gain at the BS and IRS.~In fact,~signal attenuation due to the long distance can be mitigated by applying both MISO technology and IRS.\\
	\textit{iii) IRS-aided SWIPT \textcolor{blue}{systems}:} Several works in the literature have focused on applying an IRS in a SWIPT network to improve the system efficiency\cite{20}--[30].~For instance,~the authors in \cite{20} studied the weighted sum-power optimization problem in an IRS-aided SWIPT network such that the wireless coverage and the battery life of devices in the IRS-assisted SWIPT network were extended.~In particular,~it generalized the result in \cite{21} by showing that for users with any arbitrary channels,~dedicated energy beamforming is not required,~which thus helps lower the implementation complexity.~In \cite{22},~minimizing the total transmission power was investigated in the MISO IRS-aided SWIPT networks,~where a penalty-based algorithm was proposed.~Also,~\cite{23} examined max-min received energy at the EHRs where alternating algorithm (AO) and \textcolor{blue}{semi-definite} relaxation (SDR) techniques were applied to address the optimization problem suboptimally.~Besides,~the multiple-input multiple-output (MIMO) scenario in an IRS-aided network was considered in \cite{24},~where the weighted sum data rate of the IDRs was maximized.~Furthermore,~an iterative algorithm with the lower complexity was proposed to strike a balance between a optimality and computational complexity. \cite{Zargari} studied a MISO IRS-assisted SWIPT system in which the maximization of EE indicator is introduced to establish a trade-off between the data rate and EH via joint active and passive beamforming at the BS and IRS, respectively, as well as PS ratio at each user.
	\subsection{Motivation and Contributions}
	There are appealing advantages regarding IRS systems: First, an IRS consumes much less power compared with traditional relays and naturally operates in the full-duplex mode without the need for performing any self-interference cancellation (SI). Second, both the direct-path and reflect-path in an IRS system carrying useful information data and can be coherently combined at the receiver side to improve the total received power. As a result, IRS embodies all of the qualities to facilitate the deployment of energy-efficient communication systems. Motivated by the aforementioned observations, we focus on the IRS performance along with the non-linear PS-based SWIPT technique, including co-located receivers from the max-min EE point of view (see Table I). Our contributions are summarized as follows\textcolor{blue}{:}
	
	\begin{enumerate}

		\item[$\bullet$] Compared with \cite{20}--[27],~where the authors only investigated the system performance of IRS-assisted SWIPT systems from the aspect of separated receiving nodes either as EHRs or IDRs, we consider co-located receivers based on the PS architecture. Also, we consider the optimization of the energy beam at the transmitter side in contrast to \cite{20,22,24,Zargari}, where no dedicated energy beam was considered.

		\item[$\bullet$]  In contrast to previous studies, e.g., \cite{20}--[28], investigated the resource allocation design adopting an only simplified linear EH model, we consider non-linear EH receivers in an IRS-aided SWIPT system which leads to a fundamentally different optimization problem.

		\item[$\bullet$] We take into account resource allocation fairness for all users. To this end, a max-min EE problem with the joint design of transmission information/energy beams at the BS, PS ratio at each user, and phase shifts at the IRS are investigated considering a minimum required data rate as well as energy, transmission power, and phase shifts constraints.

		\item[$\bullet$] In general,~the max-min EE optimization problem is a non-convex fractional program.~To tackle this,~we first apply the SDR, successive convex approximation (SCA) techniques, and then the difference of concave functions (DC) programming by applying the majorization-minimization (MM) approach to establish a concave-convex function.~The numerator of the non-linear objective function is first approximated by adopting the sequential fractional programming (SFP) approach \cite{bjorson}, and then the Dinkelbach method is applied to transform the obtained objective function into an equivalent subtractive. Finally, the tightness of the approximate solution is revealed.
		
		\item[$\bullet$]  Two iterative algorithms based on the penalty method and IA are proposed to handle the unit modulus constraints. For the penalty-based algorithm, information/energy beams at the BS and PS ratios are optimized in the first sub-problem. Then, phase shifts at the IRS are designed in the second sub-problem. While for the IA-based algorithm, PS ratios are optimized in the first sub-problem, and the information/energy beams at the BS and phase shifts at the IRS are optimized jointly in the second sub-problem. Finally, we propose an efficient approach based on the AO iterative algorithm to solve corresponding sub-problems in each algorithm and achieve an efficient suboptimal solution.
		
		\item[$\bullet$] Numerical results show that the EE can be significantly improved by the deployment of the IRS as compared with the conventional benchmarks with random phase shifts at the IRS and fixed PS ratios.	
	\end{enumerate}	 
	
	\textit{Notation:} Vectors and matrices are expressed by
	boldface lower case letters $\mathbf{a}$ and capital letters $\mathbf{A}$,~respectively.~For a square matrix $\mathbf{A}$,~$\mathbf{A}^H$,~$\mathbf{A}^T$,
	$\text{Tr}(\mathbf{A})$,~$||\mathbf{A}||_{*}$,~and $\text{Rank}(\mathbf{A})$ are Hermitian conjugate transpose,~transpose,~trace,~trace norm of matrix, and rank of a matrix,~respectively.~$\mathbf{A}\succeq\mathbf{0}$ indicates a positive \textcolor{blue}{semi-definite} matrix.~$\mathbf{I}_M$ denotes the $M$-by-$M$ identity matrix.~$\text{diag}(\cdot)$ is the diagonalization operation.~Diag ($\bold{A}$) indicates a vector whose
	elements are extracted from the main diagonal elements of matrix $\mathbf{A}$. The Euclidean norm of a complex vector and the absolute value of a complex scalar are denoted by $\|\cdot\|$ and $|\cdot|$,~respectively.~The distribution of a circularly symmetric complex Gaussian (CSCG) random vector with mean $\boldsymbol{\mu}$ and covariance matrix $\mathbf{C}$ is denoted by $\sim \mathcal{C}\mathcal{N}(\boldsymbol{\mu},\,\mathbf{C})$.~$\nabla_{\mathbf{x}}$ denotes the gradient vector with respect to $\mathbf{x}$. The expectation operator is denoted by $\mathbb{E}[\cdot]$,~and $\mathbb{C}^{M\times N}$ represents $M\times N$ dimensional complex matrices.~$\mathcal{O}$ expresses the big-O notation \cite{Bigo}.
	
	\section{System Model and Problem Formulation}
	\subsection{System Model}
	In this paper,~we consider a downlink (DL) MISO IRS-aided SWIPT system.~As shown in Fig.~1,~a multi-antenna BS transmits signals to $K$ single-antenna users denoted by $\mathcal{K}=\{1,...,K\}$.~The users receive the intended signals from the reflected path,~where the reflected path originates from the IRS.~We consider a reflecting IRS, which is designed based on a digital coding reflective metasurface. The elements in the metasurface contain varactor diodes with a tunable biasing voltage \cite{B5G}.~Moreover,~the BS and the IRS are equipped with $M$ transmitting antennas and $N$ reflecting elements,~respectively,~and an IRS controller is also assumed to acquire the channel state information (CSI) perfectly\footnote{In general, with the existence of RF chains at the IRS, traditional channel estimation techniques can be used to estimate the channels links of the AP-IRS and IRS-user, respectively. Without the presence of RF chains, uplink pilots in conjunction with IRS reflection patterns can be designed to estimate the corresponding channel links \cite{77,16}. Besides, the signaling overhead is the information (some complex numbers) needed to be exchanged between the BS and the IRS.}.~In particular,~energy signal is considered at the BS to facilitate potential EH at the users.~Thus,~the transmitted signal can be represented as
	\begin{equation}\label{transmit signal}
	\mathbf{x} = \underbrace {\sum\limits_{k = 1}^K {{\mathbf{w}_k}{s_k}} }_{\text{desired}\:{\rm{ }}\text{information}{\rm{ }}\:\text{signals}} + \underbrace { {{\mathbf{w}_{\mathrm{E}}}} }_{\text{energy}\:{\rm{ }}\text{signal}},\\
	\end{equation}
	where $s_{k}$ is the intended data for user $k$ and is assumed to follow the Gaussian distribution i.e.,~$s_{k} \sim \mathcal{C}\mathcal{N}(0,\,1),~\forall k$.~${\mathbf{w}}_{k}\in \mathbb{C}^{M \times 1}$ is the corresponding transmit beamforming vector.~$\mathbf{w}_{\mathrm{E}}$ is an energy signal generated at the transmitter,~which is a Gaussian pseudo-random sequence and can be exploited by users to harvest energy,~i.e.,~$\mathbf{w}_{\mathrm{E}} \sim \mathcal{C}\mathcal{N}(0,\,\mathbf{W}_{\mathrm{E}})$,~where $\mathbf{W}_{\mathrm{E}}$ denotes the covariance matrix of energy signal at the transmitter.~Moreover,~a quasi-static flat fading channel model is considered for all channels. The baseband equivalent channel responses from the BS-to-IRS,~IRS-to-user $k$, and BS-to-user $k$ are denoted by ${\mathbf{H}}\in \mathbb{C}^{N \times M}$,~${\mathbf{h}}_{ru,k}\in \mathbb{C}^{N\times 1}$,~and $\mathbf{ h}_{bu,k}\in \mathbb{C}^{M\times 1}$,~respectively.~The reflection-coefficients matrix adopted at the IRS is given by ${\mathbf{\Theta}}=\text{diag} (\beta_{1} e^{j\alpha_{1}},~\beta_{2} e^{j\alpha_{2}},...,~\beta_{N} e^{j\alpha_{N}})$,~where $ \alpha_{n} \in (0,2\pi]$ and $\beta_{n} \in [0,1]$,~$\forall n \in \{1,...,N\}$,~are phase shift and amplitude of the $n$-th element,~respectively.~For ease of implementation in practice,~it is assumed that the reflection amplitudes of all elements have the maximum values\footnote{Further control of the amplitude of each reflecting component can offer a higher adaptability and flexibility in reshaping the reflected signals \cite{amplitud,Guo}, which is left for future work.},~i.e.,~$\beta_{n}=1,~\forall n$ \cite{16}. Besides, according to the measurements presented in \cite{TangTang}--[50], the reflection amplitude/efficiency of the IRS can approach one.~Due to the high path loss,~it is assumed that the power of the signals
	that are reflected by the IRS two or more times is negligible and thus ignored\cite{16}.~The received signal at user $k$ can be written as
	\begin{equation}\label{received}
	{y_k} = {\mathbf{h}}_k^H{\mathbf{x}} + {n_k}\textcolor{blue}{,~\forall k,}
	\end{equation}
	where $n_{k}\sim \mathcal{C}\mathcal{N}(0,\,\sigma_{k}^2)$ is the antenna noise at user $k$ with zero mean
	and variance $\sigma_{k}^2$.~In particular,~the equivalent channel to user $k$,~i.e.,~${\mathbf{h}}_{k}$,~is defined as			\begin{figure}[t]
		\centering
		\includegraphics[width=3.5in]{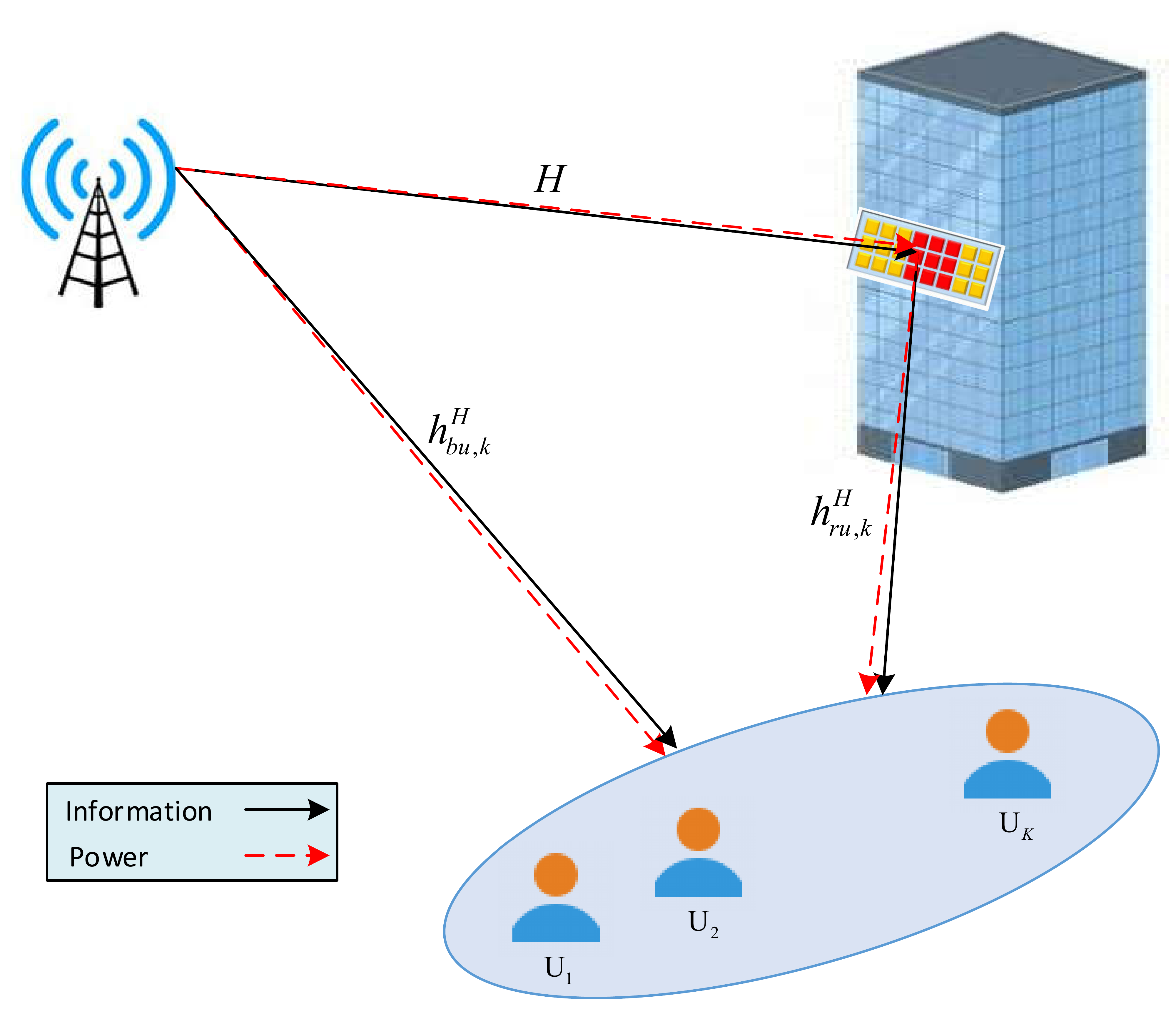}
		\caption{\small A multi-user MISO IRS-aided SWIPT system.}
	\end{figure} 
	\begin{equation}\label{equivalent}
		{\mathbf{h}}_k^H \buildrel \Delta \over = \underbrace {{\mathbf{h}}_{ru,k}^H{\mathbf{\Theta}} {\mathbf{{\mathbf{H}}}}}_{\text{reflected path}}+ \underbrace {\mathbf{ h}_{bu,k}^H}_{\text{direct path}}\textcolor{blue}{,~\forall k,}
		\end{equation}
	Furthermore,~a PS-based receiver architecture is considered at each user in which $0\leq\rho_{k}\leq 1$ is the portion of the received signal used for ID and the remaining $1-\rho_{k}$ portion is for EH.~Consequently,~the received signal at user $k$ for ID can be expressed as
	\begin{equation}\label{ID}
	y_{k}^{\text{ID}}=\sqrt{\rho_{k}}({\mathbf{{\rm{ {\mathbf{h}}}}}}_{k}^H {\mathbf{x}} + {n_k})+z_{k}\textcolor{blue}{,~\forall k,}
	\end{equation}
	where  $z_{k}\sim \mathcal{C}\mathcal{N}(0,\,\delta_{k}^2)$ is the additional noise due to the base-band signal processing at the ID receiver with zero mean and variance $\delta_{k}^2$ \cite{Liu}.~By assuming that the interference with respect to $\mathbf{w}_{\mathrm{E}}$ can be eliminated perfectly at each receiver for ID \cite{33},~the SINR at user $k$ is given by
	\begin{equation}\label{sinr}
	\text{SINR}_k= \frac{{{\rho _k}{{\left| {\mathbf{{\rm{ {\mathbf{h}}}}}}_{k}^H {\mathbf{{{w}}}}_k \right|}^2}}}{{{\rho _k}\sum\limits_{\scriptstyle i= 1\atop\scriptstyle i \ne k}^K {{{\left|  {\mathbf{{\rm{ {\mathbf{h}}}}}}_{k}^H {\mathbf{{{w}}}}_i \right|}^2} + {\rho _k}\sigma _k^2 + \delta _k^2} }},~\forall k.\\
	\end{equation}
	On the other hand,~the received signal at DL user $k$ for EH can be written as
	\begin{equation}\label{EH}
	y_{k}^{\text{EH}}=\sqrt{1-\rho_{k}}( {\mathbf{{\rm{ {\mathbf{h}}}}}}_{k}^H{\mathbf{x}} + {n_k})\textcolor{blue}{,~\forall k.}
	\end{equation}
	Accordingly,~the splitted power for EH at each user is
	\begin{equation}\label{harvested power}
	{P_k}=(1-\rho_{k})\mathbb{E}\bigg\{\big(\sum\limits_{i = 1}^K {{\left|  {\mathbf{{\rm{ {\mathbf{h}}}}}}_{k}^H {\mathbf{{{w}}}}_i \right|}^2}+{{\left|  {\mathbf{{\rm{ {\mathbf{h}}}}}}_{k}^H {\mathbf{w}_{\mathrm{E}}} \right|}^2}\big)\bigg\}\textcolor{blue}{,~\forall k.}
	\end{equation}
	Note that the power of the antenna noise,~i.e.,~$\delta^{2}_{k}$,~is ignored for EH since it is negligible. From the practical point of view,~a parametric non-linear (NL) EH model based on the sigmoidal function is introduced \cite{Nonlinear}.~Consequently,~the total harvested energy at user $k$ is modeled as
	\begin{subequations}\label{EH}\begin{align}\label{EH}
		&  \phi _k^{\text{NL}} = \frac{{\psi _k^{\text{NL}} - {M_k}{\Omega _k}}}{{1 - {\Omega _k}}},~\quad {\Omega _k} = \frac{1}{{1 + \exp ({a_k}{b_k})}}\textcolor{blue}{,~\forall k,}\\
		& \psi _k^{\text{NL}} = \frac{{{M_k}}}{{1 + \exp ( - {a_k}({P_k} - {b_k}))}}\textcolor{blue}{,~\forall k,}\label{8b}
		\end{align}
	\end{subequations}
	where $\psi _k^{\text{NL}}$ is the traditional logistic function and ${\Omega _k}$ is a constant to guarantee a zero input/output response.~Moreover,~parameters $a_{k}$ and $b_{k}$ are constants related to the circuit characteristics such as the capacitance,~resistance,~etc.~$M_{k}$ is a constant which denotes the maximum harvested power at user $k$ when the EH circuit is saturated.~Note that the parameters $a_{k}$,~$b_{k}$,~and $M_{k}$ can be obtained by a curve fitting tool\cite{Nonlinear}.~In the following,~the proposed NL EH model is adopted for formulating the EE fairness of the system and resource allocation algorithm design.~Define the EE for each user as the data rate for user $k$ per total dissipation power which is given by
	\begin{equation}
	\eta_{\mathrm{EE}}(\rho_{k},{\mathbf{w}}_k,{\mathbf{\Theta}},\mathbf{W}_\mathrm{E})=\frac{R_k(\rho_{k},{\mathbf{w}}_k,{\mathbf{\Theta}})}{{P_{T}}_k({\mathbf{w}}_k,\mathbf{W}_{\mathrm{E}})},~\: \forall k,
	\end{equation} 
	where
	\begin{align}
	&R_k(\rho_{k},{\mathbf{w}}_k,{\mathbf{\Theta}})={{\log_{2} (1 + \text{SINR}_k)} },~\forall k,\\
	&{{P_{T}}_k}({\mathbf{w}}_k,\mathbf{W}_{\mathrm{E}})=||\mathbf{w}_k||^2+\text{Tr}({\mathbf{W}}_\mathrm{E})+{P_{CR}}_{k}\textcolor{blue}{,~\forall k,}
	\end{align}
	are the data rate of each user and the total transmit power of the BS to user $k$, respectively.~Besides,~${P_{CR}}_{k}$ denotes the circuit power consumption at receiver $k$ \cite{Schober}.
	\subsection{ Problem Formulation }
	In this section,~we aim at maximizing the minimum EE of users to guarantee fairness among them by jointly optimizing the PS ratios,~transmit beamforming vectors at the BS,~and phase shifts at the IRS.~Accordingly,~the optimization problem can be formulated as
	\begin{subequations}
		\begin{align}
		& \text{(P1)}:  \underset{\rho_{k},{\mathbf{w}}_k,{\mathbf{\Theta}},\mathbf{W}_{\mathrm{E}}} {\text{maximize}} \: \underset{k}{\text{min}}\:\: \left \{\eta_{\mathrm{EE}}(\rho_{k},{\mathbf{w}}_k,{\mathbf{\Theta}},\mathbf{W}_\mathrm{E})\right\} \\
		&\text{s.t.} \quad \frac{{{\rho _k}{{\left| {{{\rm{ {\mathbf{h}}}}}}_{k}^H {\mathbf{{{w}}}}_k \right|}^2}}}{{{\rho _k}\sum\limits_{\scriptstyle i= 1\atop\scriptstyle i \ne k}^K {{{\left|  {{{\rm{ {\mathbf{h}}}}}}_{k}^H {\mathbf{{{w}}}}_i \right|}^2} + {\rho _k}\sigma _k^2 + \delta _k^2} }}\geq \gamma_{k},\: \forall k,\label{11b}\\
		&\quad\quad  \phi _k^{\text{NL}}\geq \text{E}_{\text{min},k},~\: \forall k,\label{11c}\\
		&\quad\quad \sum\limits_{k = 1}^K {{{\left\| {\mathbf{{{w}}}}_k \right\|}^2}}+\text{Tr}({\mathbf{W}}_\mathrm{E}) \le {p_{\max }},\label{11d}\\
		&\quad\quad 0 < \rho_{k} < 1,~\: \forall k,\label{11e}\\
		&\quad\quad  \mathbf{W}_{\mathrm{E}}\succeq \mathbf{0},\label{11f}\\
		&\quad\quad  {{|\mathbf{\Theta}_{{nn}}|}=1,~\forall n.\label{11g}}
		\end{align}
	\end{subequations}
	In (P1),~$\gamma_{k}$,~$\text{E}_{\text{min},k}$,~and $p_{\text{max}}$ are the target SINR,~minimum required harvested power for user $k$,~and the maximum transmit power of the BS,~respectively.~Constraint (\ref{11b}) is imposed to guarantee the quality-of-service (QoS) requirement of each user in term of received SINR.~Constraint (\ref{11c}) guarantees the amount of harvested power at each user.~Constraint (\ref{11d}) limits the
	total transmit power of the BS.~(\ref{11e}) is the PS ratio constraint at each user.~(\ref{11f}) is the positive \textcolor{blue}{semi-definite} Hermitian matrix regarding the energy signal.~(\ref{11g}) ensures that the diagonal phase shift matrix has $N$ unit modulus components on its main diagonal.~However,~the constraints set of problem (P1) is a non-convex one,~which renders the problem intractable.~To address this issue,~we first express these constraints in their equivalent forms. In particular,~we deal with the non-convex constraints (\ref{11b}) and (\ref{11c}) related to the QoS requirement and EH provisioning for solving problem (P1).~In particular,~constraint (\ref{11b}) can be equivalently stated as  
	\begin{equation}\label{17}
	\frac{{{{\left| {{\mathbf{h}}_k^H{{\mathbf{w}}_k}} \right|}^2}}}{{{\gamma _k}}}-\sum\limits_{\scriptstyle i = 1\atop\scriptstyle i \ne k}^K {{{\left|  {{\mathbf{h}}_k^H{{\mathbf{w}}_i}} \right|}^2}}  \ge \sigma _k^2 + \frac{{\delta _k^2}}{{{\rho _k}}}\textcolor{blue}{,~\forall k,}
	\end{equation}
	where (\ref{17}) is a convex constraint with respect to $\rho_{k}$ due to the fact that $\frac{1}{\rho_{k}}$ is a convex function.~Note that since ${{\Omega}}_k$ does not have any effect on the design of the optimization problem,~so $\psi _k^{\text{NL}}$ is directly employed to describe the harvested power at the $k$-th cellular user.~Hence,~the inverse function of (\ref{8b}),~i.e.,~$\psi _k^{\text{NL}}$ can be written as
	\begin{equation}\label{14}
	{P_k}(\psi _k^{\text{NL}}) = {b_k} - \frac{1}{{{a_k}}}\ln (\frac{{{M_k} - \psi _k^{\text{NL}}}}{{\psi _k^{\text{NL}}}})\textcolor{blue}{,~\forall k.}
	\end{equation}
	By utilizing (\ref{14}),~constraint (\ref{11c}) can be transformed into
	\begin{equation}\label{equi 19}
	\sum\limits_{i = 1}^K {{\left|  {\mathbf{{\rm{ {\mathbf{h}}}}}}_{k}^H {\mathbf{{{w}}}}_i \right|}^2}+{{\left|  {\mathbf{{\rm{ {\mathbf{h}}}}}}_{k}^H {\mathbf{w}_{\mathrm{E}}} \right|}^2} \ge \frac{{{P_k}({\text{E}_{\text{min},k}})}}{{1 - {\rho _k}}}\textcolor{blue}{,~\forall k,}
	\end{equation}
	which is a convex constraint with respect to $\rho_{k}$ since $\frac{1}{{1-\rho_{k}}}$ is a convex function.
	\section{Proposed Algorithms}
In the following,~we first adopt the AO algorithm for two proposed algorithms to handle problem (P1) efficiently.
	\subsection{Penalty-Based Algorithm}
	In penalty-based, for the first sub-problem,~the semi-definite relaxation (SDR) approach is applied to optimize the beamforming vectors \textcolor{blue}{at the BS} and PS ratios jointly.~Then,~we rewrite the data rate in a class of difference of concave functions  (DC) and we employ successive convex approximation (SCA) to make a convex data rate.~Then,~we transform the optimization problem into a smooth one by introducing an auxiliary variable.~However,~the obtained problem is in a fractional form,~which needs to be decomposed into an equivalent linear objective function by exploiting the Dinkelbach method.~While applying the similar SDR method to the second sub-problem,~the phase shifts are optimized for the given beamforming vectors and PS ratios based on the penalty function.~The key steps for obtaining the solution of the considered optimization problem are given in Fig.~2.
	
	\subsubsection{Sub-problem 1: Joint Beamforming Design and PS Ratio with Fixed Phase Shifts}
	By applying the AO algorithm with fixed phase shifts, i.e., ${\mathbf{\Theta}}$, problem (P1) can be reformulated as the following equivalent problem\textcolor{blue}{:}
	\begin{subequations}
		\begin{align}
		& \text{(P2)}:  \underset{\rho_{k},{\mathbf{w}}_k,\mathbf{W}_{\mathrm{E}}} {\text{maximize}} \: \underset{k}{\text{min}}\:\:\left \{\frac{{R}_{k}({\mathbf{w}}_k,\rho_{k})}{{{P_{T}}_k}({\mathbf{w}}_k,\mathbf{W}_{\mathrm{E}})}\right \}\\
		&\text{s.t.}\quad \text{(\ref{11b})--(\ref{11f})}.
		\end{align}
	\end{subequations}
	By applying the SDR technique,~problem (P2) can be handled efficiently.~Define ${\mathbf{{W}}}_{k}={\mathbf{w}}_{k}{\mathbf{w}}_{k}^H$,~$\forall k$,~where matrix ${\mathbf{W}}_{k}$ is semi-definite and satisfies  $\text{Rank}({\mathbf{{W}}}_{k})\leq1$,~$\forall k$,~problem (P2) can be restated as
	\begin{subequations}
		\begin{align}
		&\text{(P3)}:  \underset{\rho_{k},{\mathbf{W}}_k,\mathbf{W}_{\mathrm{E}}} {\text{maximize}} \: \underset{k}{\text{min}}\:\:\left \{\frac{{R}_{k}({\mathbf{W}}_k,\rho_{k})}{{{P_{T}}_k}({\mathbf{W}}_k,\mathbf{W}_{\mathrm{E}})}\right \}\\
		&\text {s.t.} \quad \frac{{\text{Tr}({{\mathbf{H}}_k}{{\mathbf{W}}_k})}}{{{\gamma _k}}} - \sum\limits_{\scriptstyle i = 1\atop\scriptstyle i \ne k}^K {\text{Tr}({{\mathbf{H}}_k}{{\mathbf{W}}_i})}  \ge \sigma _k^2 + \frac{{\delta _k^2}}{{{\rho _k}}},\: \forall k,\label{22c}\\
		&\quad\quad \sum\limits_{i = 1}^K {{{ \text{Tr}({{\mathbf{H}}_k}{{\mathbf{W}}_i}) }}}+\text{Tr}({{\mathbf{H}}_k}\mathbf{W}_\mathrm{E})  \ge \frac{{{P_k}({\text{E}_{\text{min},k}})}}{{1 - {\rho _k}}},~\: \forall k,\label{22d}\\
		&\quad\quad \sum\limits_{k = 1}^K {\text{Tr}({{\mathbf{W}}_k})}+{\text{Tr}({{\mathbf{W}}_{\mathrm{E}}})}  \le p_\text{max},\label{power}\\
		&\quad\quad 0 < \rho_{k} < 1,~\: \forall k,\\
		&\quad\quad {{\mathbf{W}}}_{k},{{\mathbf{W}}}_{\mathrm{E}} \succeq
		\mathbf{0},~\: \forall k,\label{22g}\\
		&\quad\quad\text{Rank}({\mathbf{{W}}}_{k})\leq1,~\forall k,~   \label{22h} 
		\end{align}
	\end{subequations}
	where 
	\begin{align}
	&{R}_k({\mathbf{W}}_k,\rho_{k})=\text{log}_2\bigg({{{ {\text{Tr}({{\mathbf{H}}_k}{{\mathbf{W}}_k})}}}}+\sum\limits_{\scriptstyle i= 1\atop\scriptstyle i \ne k}^K {{ {\text{Tr}({{\mathbf{H}}_k}{{\mathbf{W}}_i})}}} +\sigma _k^2 \nonumber\\ &+\frac{\delta _k^2}{{\rho _k}}\bigg)-\text{log}_2\bigg({\sum\limits_{\scriptstyle i= 1\atop\scriptstyle i \ne k}^K {{{  {\text{Tr}({{\mathbf{H}}_k}{{\mathbf{W}}_i})} }} + \sigma _k^2 +\frac{\delta _k^2}{{\rho _k}} } }\bigg)\textcolor{blue}{,~\forall k,}\label{20}\\
	&{{P_{T}}_k}({\mathbf{W}}_k,\mathbf{W}_{\mathrm{E}})=\text{Tr}({\mathbf{W}}_k)+ \text{Tr}(\mathbf{W}_\mathrm{E})+ {P_{CR}}_{k}\textcolor{blue}{,~\forall k.}
	\end{align}
	Since (\ref{20}) belongs to the class of DC programming,~it can be written as ${R}_k(\rho_{k},{\mathbf{W}}_k)=f(\mathbf{W}_k,\rho_{k})-g(\mathbf{W}_k,\rho_{k})$,~$\textcolor{blue}{\forall k,}$ where
	\begin{align}
	&{\small f(\mathbf{W}_k,\rho_{k})=\text{log}_2\bigg({\sum\limits_{\scriptstyle i= 1}^K {{{ {\text{Tr}({{\mathbf{H}}_k}{{\mathbf{W}}_i})}}} +\sigma _k^2 +\frac{\delta _k^2}{{\rho _k}}} }\bigg),}\label{200}\\&
	g(\mathbf{W}_k,\rho_{k})=\text{log}_2\bigg({\sum_{ i \ne k}^K {{{  {\text{Tr}({{\mathbf{H}}_k}{{\mathbf{W}}_i})} }} + \sigma _k^2 +\frac{\delta _k^2}{{\rho _k}} } }\bigg)\label{210}.
	\end{align}
	In this way,~$\text{(P3)}$ can be reformulated as a classic DC problem where a locally optimal solution can be obtained by the SCA method \cite{TWC_Ata,Ata_WCNC,Ata_ICC}. Since,~$g(\mathbf{W}_k,\rho_{k})$ is a differentiable convex function with respect to both $\mathbf{W}_k$ and $\rho_k$,~we have
	{\small \begin{align}\label{26}
	&g(\mathbf{W}_k,\rho_{k})\leq g(\mathbf{W}_k^{(t-1)},\rho_{k}^{(t-1)})+\text{Tr}\bigg(\big(\nabla_{\mathbf{W}_k}^Hg(\mathbf{W}_k^{(t-1)},\rho_k^{(t-1)})\big)\nonumber\\ &(\mathbf{W}_k-\mathbf{W}_k^{(t-1)})\bigg)+\text{Tr}\bigg(\big(\partial _{\rho_k}g(\mathbf{W}_k^{(t-1)},\rho_k^{(t-1)})\big)\big(\rho_k-\rho_k^{(t-1)}\big)\bigg)\nonumber\\ &\triangleq\tilde{g}(\mathbf{W}_k,\rho_k),
	\end{align}}
	where
	\begin{align}
	&\nabla_{\mathbf{W}_k}^Hg(\mathbf{W}_k^{(t-1)},\rho_k^{(t-1)})=\frac{\mathbf{e}_k}{{\sum\limits_{\scriptstyle \atop\scriptstyle i \ne k} {{{  {\text{Tr}({{\mathbf{H}}_k}{{\mathbf{W}}_i^{(t-1)}})} }} + \sigma _k^2 +\frac{\delta _k^2}{{\rho^{(t-1)}_k}} } }},\label{27}\\
	&\partial _{\rho_k}g(\mathbf{W}_k^{(t-1)},\rho_k^{(t-1)})=-\frac{\delta^2_{k}/\rho^{2,{(t-1)}}_k}{\ln2{\sum\limits_{\scriptstyle \atop\scriptstyle i \ne k} {{{  {\text{Tr}({{\mathbf{H}}_k}{{\mathbf{W}}_i^{(t-1)}})} }} + \sigma _k^2 +\frac{\delta _k^2}{{\rho^{(t-1)}_k}} } }}.\label{277}
	\end{align}\begin{figure}
		\centering
		\includegraphics[width=3.5in]{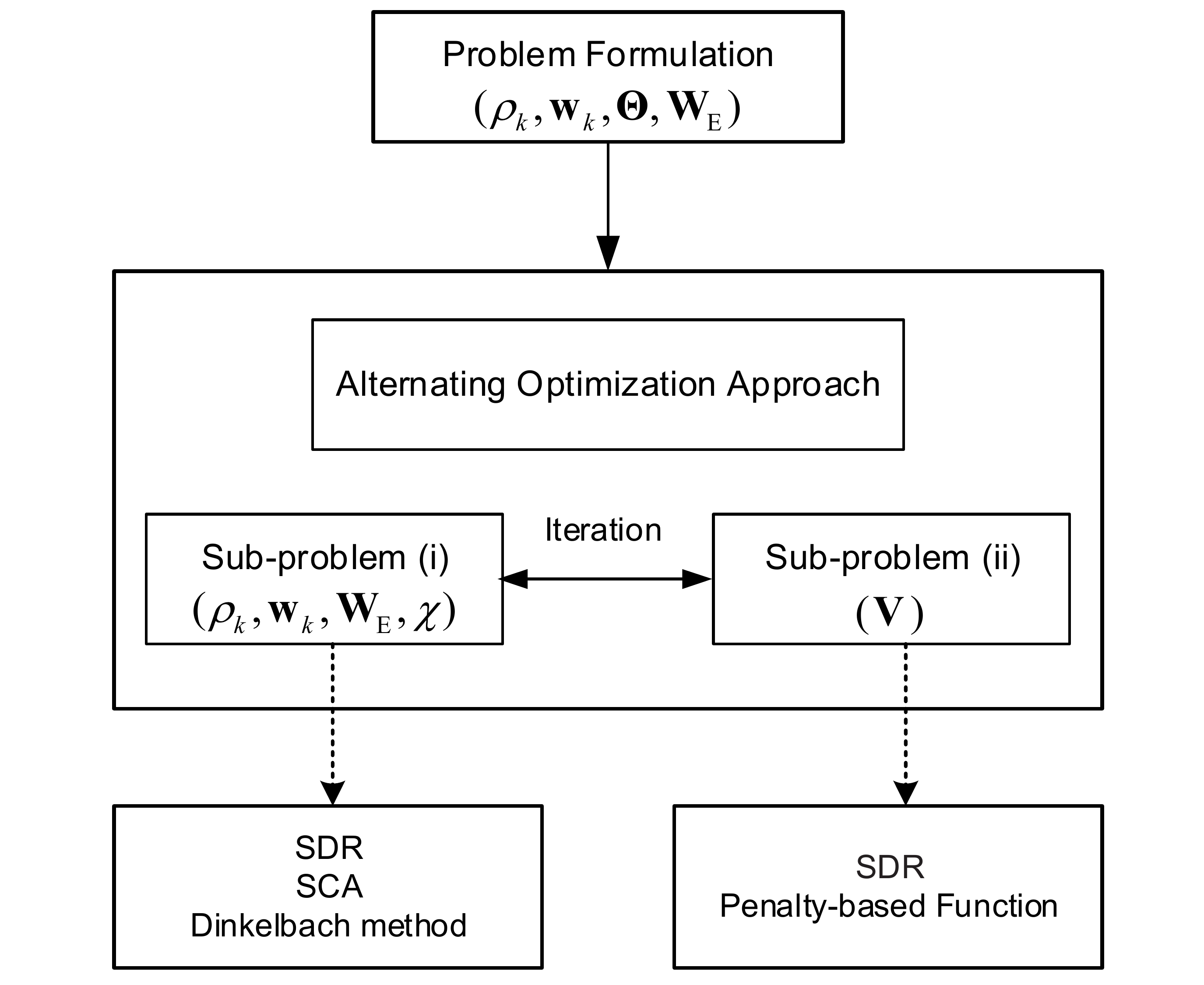}
		\caption{\small A flow chart of the proposed penalty-based algorithm.}
	\end{figure}
\begin{algorithm}[t]
	\caption{Iterative Resource Allocation Algorithm}
	\begin{algorithmic}[1]
		\renewcommand{\algorithmicrequire}{\textbf{Input:}}
		\renewcommand{\algorithmicensure}{\textbf{Output:}}
		\REQUIRE Set maximum number of iterations $I_\text{max}=10$ and the
		maximum error tolerance $\epsilon=10^{-2}$.~Set initial iteration $i=0$ and the maximum energy efficiency $\lambda^{(i)}$=0.\\    
		\STATE \textbf{repeat}\\
		\STATE \quad Solve problem (P6) for a given $\lambda$ and
		obtain\\ \quad $\{\rho_{k}^{(i)},\chi^{(i)},\mathbf{w}_k^{(i)},\mathbf{W}_{\mathrm{E}}^{(i)}\}.$\\
		\STATE \quad \textbf{if} ${\small \big|\underset{k}{\text{min}}\:\: \left\{{\tilde{R}_{k}({\mathbf{W}}^{(i)}_k,\rho_{k}^{(i)})}-\lambda^{(i)}{{{{P_{T}}_k}({\mathbf{W}}_k^{(i)},\mathbf{W}_{\mathrm{E}}^{(i)})}}\right\}\big|\leq \epsilon}  $\\
		\STATE \quad \quad	 \textbf{then} \\
		\STATE \quad \quad \textbf{return} $\{\rho_{k}^\ast,\chi^*,{\mathbf{w}}_k^\ast,\mathbf{W}_{\mathrm{E}}^\ast\}=\{\rho_{k}^{(i)},\chi^{(i)},{\mathbf{w}}_k^{(i)},\mathbf{W}_{\mathrm{E}}^{(i)}\}$\\ \quad \quad and $\lambda^\ast=\underset{k}{\text{min}}\:\:\frac{\tilde{R}_{k}({\mathbf{W}}^{(i)}_k,\rho_{k}^{(i)})}{{{P_{T}}_k}({\mathbf{W}}_k^{(i)},\mathbf{W}_{\mathrm{E}}^{(i)})}$.\\
		\STATE \quad  \textbf{else}
		\STATE \quad \quad Set $\lambda^{(i+1)}=\underset{k}{\text{min}}\:\:\frac{\tilde{R}_{k}({\mathbf{W}}^{(i)}_k,\rho_{k}^{(i)})}{{{P_{T}}_k}({\mathbf{W}}_k^{(i)},\mathbf{W}_{\mathrm{E}}^{(i)})}$.
		\STATE \quad  \textbf{end if}
		\STATE   \textbf{until} $i=I_{\text{max}}$.
	\end{algorithmic}
\end{algorithm}	In ({\ref{27}}),~$\mathbf{e}_k$ is an $M$-dimensional column vector which can be expressed as $\mathbf{e}_k(k)= 0$ and $\mathbf{e}_k(i)=\frac{\mathbf{H}_k}{\ln2},~i\neq k$.~Motivated by SDR technique,~(P3) can be relaxed by dropping the non-convex rank-one constraint (\ref{22h}) and a locally optimal solution can be achieved as follows:
	\begin{subequations}
		\begin{align}
		&\text{(P4)}:  \underset{\rho_{k},{\mathbf{W}}_k,\mathbf{W}_{\mathrm{E}}} {\text{maximize}} \: \underset{k}{\text{min}}\:\:\left \{\frac{{\tilde{R}}_{k}({\mathbf{W}}_k,\rho_{k})}{{{P_{T}}_k}({\mathbf{W}}_k,\mathbf{W}_{\mathrm{E}})}\right \}\\
		&\text {s.t.} \quad \text{(\ref{22c})--(\ref{22g})},
		\end{align}
	\end{subequations}
	where $\tilde{R}_{k}({\mathbf{W}}_k,\rho_{k})=f(\mathbf{W}_k,\rho_{k})-\tilde{g}(\mathbf{W}_k,\rho_{k})$,~$\textcolor{blue}{\forall k}$.~The proposed algorithm for solving (P4) \textcolor{blue}{is given in Algorithm 1, where} the objective function of (P4) at iteration $t$ is $f(\mathbf{W}_k,\rho_{k})-\tilde{g}(\mathbf{W}_k,\rho_{k})$. Consequently, we have the following \textcolor{blue}{proposition}:
	\begin{proposition}
			The approximation (\ref{26}) produces a tight lower bound of ${R}_{k}({\mathbf{W}}_k,\rho_{k})$ which leads to a sequence of improved solutions for (P4).
	\end{proposition}
	
	\textit{Proof:}
		The proof is provided in Appendix A.\quad\quad\quad\quad\quad $\blacksquare$
	\subsubsection{Transformation of the Objective Function}
	In this subsection,~we apply NL fractional programming theory \cite{ Dinkelbach} to solve sub-problem 1.~In particular,~the fractional problem has the following form:
	\begin{equation}\label{22}
	\lambda^\ast=\underset{\rho_{k},{\mathbf{W}}_k,\mathbf{W}_{\mathrm{E}}} {\text{maximize}} \: \underset{k}{\text{min}}\:\: \left\{{\frac{\tilde{R}_{k}({\mathbf{W}}_k,\rho_{k})}{{{P_{T}}_k}({\mathbf{W}}_k,\mathbf{W}_{\mathrm{E}})}}\right\},
	\end{equation}
	where (\ref{22}) can be transformed into an equivalent form by employing the following theorem.
	\begin{theorem}
		The fractional problem (P4) has an equivalent subtractive form which satisfies\textcolor{blue}{:}
		\begin{align}\label{13}
		&\underset{\rho_{k},{\mathbf{W}}_k,\mathbf{W}_{\mathrm{E}}} {\text{maximize}} \: \underset{k}{\text{min}}\:\: \left \{{\tilde{R}_{k}({\mathbf{W}}_k,\rho_{k})}-\lambda^\ast{{{{P_{T}}_k}({\mathbf{W}}_k,\mathbf{W}_{\mathrm{E}})}}\right\}\nonumber\\&= \underset{k}{\text{min}}\:\: \left\{{\tilde{R}_{k}({\mathbf{W}}^*_k,\rho_{k}^*)}-\lambda^\ast{{{{P_{T}}_k}({\mathbf{W}}_k^\ast,\mathbf{W}_{\mathrm{E}}^\ast)}}\right\}=0.
		\end{align}
	\end{theorem}
	\textit{Proof:}
	The proof is referred to Appendix B.\quad\quad\quad\quad\quad\quad $\blacksquare$
	
	Theorem 1 shows that (P4) can be solved by its equivalent problem (\ref{13}).
	Hence,~Dinkelbach can provide an iterative method to solve problem (\ref{13}) instead of (P4).
Finally,~the following optimization problem requires to be solved for a given $\lambda$\textcolor{blue}{:}	\begin{algorithm}[t]
 		\caption{Alternating Optimization (AO) Algorithm}
 		\begin{algorithmic}[1]
 			\renewcommand{\algorithmicrequire}{\textbf{Input:}}
 			\renewcommand{\algorithmicensure}{\textbf{Output:}}
 			\REQUIRE $j=0$,~$J_\text{max}$,~$\mathbf{\mathbf{\Theta}}^{(j)}$
 			\STATE  $j$: Number of iterations
 			\STATE  $J_\text{max}$: Maximum number of iterations
 			\STATE  $\mathbf{\mathbf{\Theta}}^{(j)}$: Initial random phases in iteration $j$\\
 			\STATE \textbf{repeat}\\
 			\STATE \quad Under given $\mathbf{\Theta}=\mathbf{\Theta}^{(j)}$,~use \textbf{Algorithm 1} to obtain\\ \quad $\{\rho_{k}^{(j)},{\mathbf{W}}_k^{(j)},\mathbf{W}_{\mathrm{E}}^{(j)},\chi^{(j)},\lambda^{(j)}\}$,~and set\\ \quad  $\mathbf{V}^{(j)}=\mathbf{v}^{(j)}(\mathbf{v}^{(j)})^H$
 			\STATE \quad  Under given optimal values of problem (P6),~solve\\ \quad problem (P9) and update $\mathbf{V}^{(j+1)}$
 			\STATE \quad  Decompose $\mathbf{V}^{(j+1)}=\mathbf{v}^{(j+1)}(\mathbf{v}^{(j+1)})^H$ and update\\ \quad $\mathbf{\Theta}^{(j+1)}=\text{diag}(\mathbf{v}^{(j+1)})$ 
 			\STATE \quad Set $j=j+1$
 			\STATE   \textbf{until} $j=J_{{\text{max}}}$.
 		\end{algorithmic}
 	\end{algorithm} 
	\begin{subequations}
		\begin{align}
		& \text{(P5)}:  \underset{\rho_{k},{\mathbf{W}}_k,\mathbf{W}_{\mathrm{E}}} {\text{maximize}} \: \underset{k}{\text{min}}\:\: \left\{{\tilde{R}_{k}({\mathbf{W}}_k,\rho_{k})}-\lambda {{P_{T}}_k}({{{\mathbf{W}}_k,\mathbf{W}_{\mathrm{E}}}})\right\} \label{14a}\\
		&\text {s.t.} \quad \text{(\ref{22c})--(\ref{22g})}.
		\end{align}
	\end{subequations}
	To facilitate the solution design and to smooth the objective function of (P5),~an auxiliary variable $\chi$ is introduced so that the equivalent form of (P5) can be expressed as\cite{Jorswieck}
	\begin{subequations}
		\begin{align}
		& \text{(P6)}:  \underset{\rho_{k},{\mathbf{W}}_k,\mathbf{W}_{\mathrm{E}},\chi} {\text{maximize}} \:\: \chi\\
		&\text{s.t.} \quad \left \{{\tilde{R}_{k}({\mathbf{W}}_k,\rho_{k})}-\lambda {{P_{T}}_k}({{{\mathbf{W}}_k,\mathbf{W}_{\mathrm{E}}}})\right\} \geq \chi,~\: \forall k,
		\label{17b}\\ 
		&\quad\quad   \text{(\ref{22c})--(\ref{22g})}.
		\end{align}
	\end{subequations}Notice that the relaxed (P6) is a standard semi-definite programming (SDP) which can be solved by using CVX \cite{CVX}.~The following proposition determines that the SDR solution provided by (P6) is tight.
	\begin{proposition}
		By denoting the optimal solutions of problem (P6) as $\{\rho_{k}^\ast,{\mathbf{W}}_k^\ast,\mathbf{W}_{\mathrm{E}}^\ast,\chi^\ast\}$,
		and assuming that all channel links are statistically independent,~then $\mathbf{W}^\ast_k$ and ${{\mathbf{W}}}^\ast_{\mathrm{E}}$  satisfy $\text{Rank}(\mathbf{W}_k)=1$ and $\text{Rank}(\mathbf{W}_\mathrm{E})\leq1$,~respectively.
	\end{proposition}

	\textit{Proof:}
	Please see Appendix C.\quad\quad\quad\quad\quad\quad\quad\quad\quad\quad\quad $\blacksquare$

	As such,~a tractable form of (P4) is obtained where we handle (P6) instead of (P4).~Next,~we attempt to optimize the phase shifts at the IRS by utilizing the AO algorithm.

	\subsubsection{Sub-problem 2: Phase Shifts Optimization with Fixed Beamforming Vectors and PS Ratios}
	In this subsection,~${\mathbf{\Theta}}$ is optimized under given optimal solutions $\{{\mathbf{W}}_{k}^\ast,{\mathbf{W}}_\mathrm{E}^\ast,\rho_k^\ast,\lambda^\ast,\chi^\ast\}$.~\textcolor{blue}{The main challenge in optimizing $\mathbf{\Theta}$ is constraint (\ref{11g}), 
	which is a module constraint and makes the problem intractable}.~Consequently,~we first define ${\boldsymbol{\theta}}=( e^{j\alpha_{1}},...,~e^{j\alpha_{N}})^H\in\mathbb{C}^{N\times1}$ and $\tilde{{\boldsymbol{\theta}}}=[{\boldsymbol{\theta}}^T \: \tau]^T\in\mathbb{C}^{(N+1)\times1}$,~respectively,~where $\tau\in\mathbb{C}$ is a dummy variable with $|\tau|=1$.~By adopting the SDP,~${\mathbf{V}}=\tilde{{\boldsymbol{\theta}}}\tilde{{\boldsymbol{\theta}}}^H\in\mathbb{C}^{(N+1)\times(N+1)}$ can be represented which follows that the matrix ${\mathbf{V}}$ is semi-definite and satisfies $\text{Rank}({\mathbf{{V}}})\leq1$. Thus, we have\begin{align}
		&|({\mathbf{h}}_{ru,k}^H{\mathbf{\Theta}} {\mathbf{{\mathbf{H}}}}+\mathbf{ h}_{bu,k}^H)\mathbf{w}_k|^2\triangleq \text{Tr}(\mathbf{V}\mathbf{L}_k\mathbf{W}_k\mathbf{L}_k^H)=\text{Tr}(\mathbf{W}_k\mathbf{Z}_k),
		\end{align}where $\mathbf{L}_k=[(\text{diag}({\mathbf{h}}_{ru,k}^H)\mathbf{H})^T\:\: \mathbf{ h}_{bu,k}^*]^T$ and $\mathbf{Z}_k=\mathbf{L}_k^H\mathbf{V}\mathbf{L}_k$.~Therefore,~(P4) can be restated as 
	\begin{subequations}
		\begin{align}
		&\text{(P7)}:  \text{Find} \quad \mathbf{V}\\
		&\text {s.t.} \quad  \text{log}_2\Big(\text{Tr}(\mathbf{W}_k\mathbf{Z}_k)+\sum\limits_{\scriptstyle i = 1\atop\scriptstyle i \ne k}^K \text{Tr}(\mathbf{W}_i\mathbf{Z}_k )+\sigma _k^2 +\frac{\delta _k^2}{{\rho _k}} \Big)\nonumber\\&\quad\quad-\tilde{g}(\mathbf{V})\geq\lambda \text{Tr}({\mathbf{W}}_k)+\lambda \text{Tr}(\mathbf{W}_\mathrm{E})+\lambda {P_{CR}}_{k}+\chi,\: \forall k,\label{28b}\\    
		&\quad\quad \frac{\text{Tr}(\mathbf{W}_k\mathbf{Z}_k)}{{\sum\limits_{\scriptstyle i = 1\atop\scriptstyle i \ne k}^K {\text{Tr}(\mathbf{W}_i\mathbf{Z}_k)+ \sigma _k^2 + \frac{{\delta _k^2}}{{\rho _k }}} }}\geq \gamma_{k},\: \forall k,\\
		&\quad\quad {\small \sum\limits_{i = 1}^K \text{Tr}(\mathbf{W}_i\mathbf{Z}_k)+\text{Tr}(\mathbf{W}_\mathrm{E}\mathbf{Z}_k)  \ge \frac{{{P_k}({\text{E}_{\text{min},k}})}}{{1 - \rho _k }},\: \forall k,}\label{28e}\\
		&\quad\quad \text{Diag}(\mathbf{V})=\mathbf{1}_{N+1},~\quad \mathbf{V}\succeq\mathbf{0},~\quad \text{Rank}(\mathbf{V})\leq 1,\label{30e}
		\end{align}
	\end{subequations}
	where we have
	\begin{align}
	&g(\mathbf{V})\leq g(\mathbf{V}^{(t)})+\text{Tr}\bigg(\nabla_{\mathbf{V}}^Hg(\mathbf{V}^{(t)})(\mathbf{V}-\mathbf{V}^{(t)})\bigg)\triangleq\tilde{g}(\mathbf{V}).
	\end{align}
	Note that (P7) is a feasibility problem and (\ref{30e}) is imposed to ensure the unit modulus constraints.~However,~due to the existence of the constraint (\ref{30e}),~(P7) usually results in a solution with a rank higher than one.~Rather than using the SDR technique,~the rank-one constraint is handled differently by exploiting the penalty term.~The equivalent form of the rank-one constraint can be expressed as
\begin{equation}\label{32}
||\mathbf{V}||_*-||\mathbf{V}||_2\leq 0.
\end{equation}
\begin{figure}[t]
	\centering
	\includegraphics[width=3in]{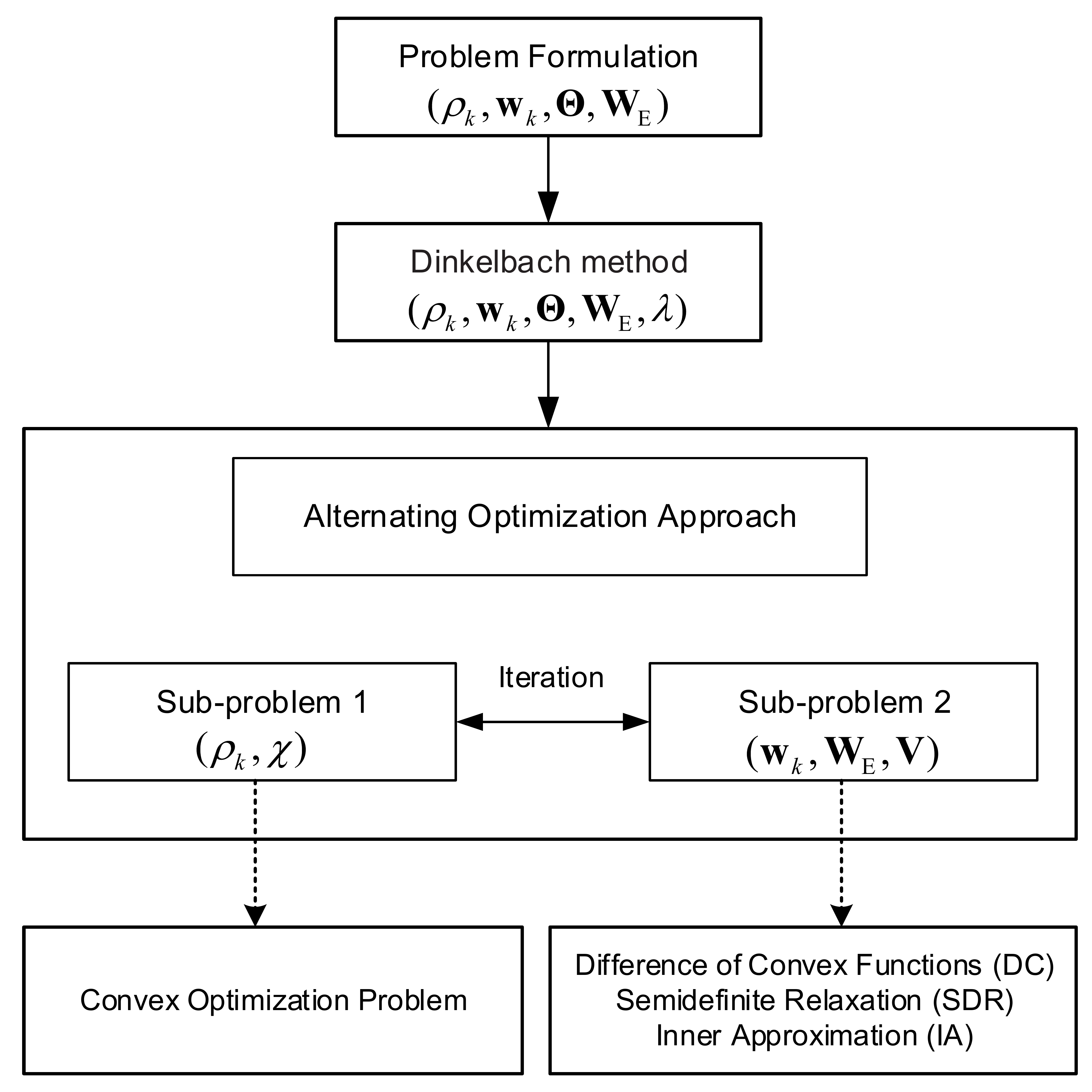}
	\caption{\small A flow chart of the proposed IA-based algorithm.}
\end{figure}Bear in mind that the inequality $||\mathbf{Y}||_*=\sum_{i} \sigma_i\geq ||\mathbf{Y}||_2=\underset{i}{\text{max}}\{\sigma_i\}$ holds for any given $\mathbf{Y}\in \mathbb{H}^{m\times n}$,~where $\sigma_{i}$ is the $i$-th singular value of $\mathbf{Y}$.~The equality holds if and only if $\mathbf{Y}$ achieves \textcolor{blue}{rank-one}.~The \textcolor{blue}{rank-one} constraint,~i.e.,~$\text{Rank}(\mathbf{V})=1$ guarantees this equality.~Hence,~we exploit the penalty-based method by integrating such a constraint into the objective function \cite{penalty},~which leads to the following optimization problem\textcolor{blue}{:}
\begin{subequations}
	\begin{align}
	& \text{(P8)}:  \underset{{\mathbf{V}}} {\text{minimize}} \:\:\frac{1}{2\mu}(||\mathbf{V}||_*-||\mathbf{V}||_2)\\
	&\text{s.t.} \quad \text{(\ref{28b})--(\ref{28e})},~\label{31b}\\
	&\quad\quad \text{Diag}(\mathbf{V})=\mathbf{1}_{N+1},~\quad \mathbf{V}\succeq\mathbf{0},\label{31c}
	\end{align}
\end{subequations}
where $\mu$ is a penalty factor for (\ref{32}) which means that the minimum value of (\ref{32}) can be obtained by applying a small $\mu$.~It is worth noting that although the \textcolor{blue}{rank-one} constraint is relaxed in (P8),~the solution achieved by solving (P8) is an optimal solution when $\mu \to 0$.~On the other hand,~for a sufficiently small value of $\mu$,~solving (P8) yields a rank-one solution.~Nevertheless,~(P8) is not convex yet due to the DC form of the objective function.~To tackle the DC form,~we define a lower bound for $A=||\mathbf{V}||_2$,~which can be obtained by
\begin{align}
{\small A(\mathbf{V})\geq A(\mathbf{V}^{(t)})+\text{Tr}\bigg(\nabla_{\mathbf{V}}^HA(\mathbf{V}^{(t)})(\mathbf{V}-\mathbf{V}^{(t)})\bigg)\triangleq\tilde{A}(\mathbf{V})},
\end{align}
Accordingly,~the optimization problem can be written as follows:
\begin{subequations}
	\begin{align}
	&\text{(P9)}:  \underset{{\mathbf{V}}} {\text{minimize}} \:\: \frac{1}{2\mu}\bigg(||\mathbf{V}||_*-\tilde{A}(\mathbf{V})\bigg)\\
	&\text {s.t.} \quad \text{ (\ref{31b})--(\ref{31c}).}
	\end{align}
\end{subequations}
This optimization problem is convex and thus can be efficiently solved by employing optimization solvers such as CVX \cite{CVX}.~The iterative SCA algorithm related to (P9) is as same as algorithm 1.~The ultimate AO algorithm is presented in Algorithm 2. 

\begin{proposition}
		(P6) is non-increasing as the objective function value increases over each iteration in Algorithm 2.~In particular, after each iteration,~the iterative algorithm in Algorithm 2 improves the objective function value of (P6).   
\end{proposition}

\textit{Proof:}
	Please see Appendix D.\quad\quad\quad\quad\quad\quad\quad\quad\quad\quad\quad $\blacksquare$

In particular, the penalty-based algorithm has a form of a two-block Gauss-Seidel algorithm, in which stationary points are achieved for both blocks determined by problems (P6) and (P9), respectively. Accordingly, the proposed algorithm converges to a stationary point of problem (P1) in polynomial time \cite{Razaviyayn}.

\subsection{IA-Based Algorithm}
In this section, another new algorithm for (P1) is investigated to optimize beamforming vectors and phase shifts at the BS and IRS, respectively, by leveraging the inner approximation (IA) method \cite{Schoberr,Green_IRS}. In this method, there is no need for penalty-based feasibility problem (P7). The idea behind this approach is that the non-convex feasible set is approximated by a convex one in each iteration, which is more tractable. Therefore, we first optimize PS ratios in one sub-problem and then apply the IA approach in the second sub-problem to optimize beamforming vectors and phase shifts, simultaneously.~The flowchart for this approach is provided in Fig.~3.

\subsubsection{Sub-problem 1: PS \textcolor{blue}{Ratios} with Fixed Phase Shifts and Beamforming Vectors}
For the given solutions in the $(t-1)$-th iteration, i.e., $\{{\mathbf{W}}_{k}^{(t-1)},{\mathbf{W}}_\mathrm{E}^{(t-1)},{\mathbf{\Theta}}^{(t-1)}\}$, the PS ratio sub-problem  is given by
\begin{subequations}
		\begin{align}
		& \text{(P10)}:  \underset{\rho_{k},\chi} {\text{maximize}} \:\: \chi\\
		&\text{s.t.} \quad R_{k}(\rho_{k})-\lambda {P_{T}}_k({\mathbf{W}}^{(t-1)}_k,\mathbf{W}^{(t-1)}_{\mathrm{E}}) \geq \chi,~\: \forall k,\\
		&\quad\quad   \text{(\ref{11e}),\:(\ref{17}),\:(\ref{equi 19})}.
		\end{align}
\end{subequations}As can be observed (P10) is a convex optimization problem which can be efficiently solved by the CVX optimization toolbox \cite{CVX}.
\subsubsection{Sub-problem 2: Phase Shifts and Beamforming Vectors with Fixed PS \textcolor{blue}{Ratios}}
	To deal with the non-convex constraints, it is straight-forward to show that $\text{Tr}(\mathbf{W}_i\mathbf{Z}_k)$ can be equivalently expressed as
\begin{align}
	\text{Tr}(\mathbf{W}_i\mathbf{Z}_k)&=\frac{1}{2}\left\|\mathbf{W}_i+\mathbf{Z}_k\right\|_F^2-\frac{1}{2}\left\|\mathbf{W}_i\right\|^2_F-\frac{1}{2}\left\|\mathbf{Z}_k\right\|^2_F\nonumber\\&\triangleq\mathbf{X}_{i,k}\textcolor{blue}{,~\forall i, k.}\label{IA}
	\end{align}which is still non-convex.~One can readily verify that equation (\ref{IA}) belongs to the class of DC function. Hence,~we resort to the majorization-minimization (MM) approach by applying the first-order Taylor approximation as follows:
\begin{align}\label{11}
	F(\mathbf{W}_i,\mathbf{Z}_k)&\triangleq\frac{1}{2}\left\|\mathbf{W}_i+\mathbf{Z}_k\right\|_F^2\geq F(\mathbf{W}^{(i)}_i,\mathbf{Z}^{(i)}_k)\nonumber\\&+\text{Tr}\left(\nabla_{\mathbf{W}_{i}}^HF(\mathbf{W}^{(i)}_i,\mathbf{Z}^{(i)}_k)(\mathbf{W}_i-\mathbf{W}^{(i)}_i)\right)\nonumber\\&+\text{Tr}\left(\nabla_{\mathbf{Z}_{k}}^HF(\mathbf{W}^{(i)}_i,\mathbf{Z}^{(i)}_k)(\mathbf{Z}_k-\mathbf{Z}^{(i)}_k)\right)\nonumber\\ &\triangleq \tilde{F}(\mathbf{W}^{(i)}_i,\mathbf{Z}^{(i)}_k)\textcolor{blue}{,~\forall i, k.}
	\end{align}{Therefore, (\ref{IA}) can be rewritten as
	{\begin{equation}
		\tilde{\mathbf{X}}^{(i)}_{i,k}\triangleq\tilde{F}(\mathbf{W}^{(i)}_i,\mathbf{Z}^{(i)}_k)-\frac{1}{2}\left\|\mathbf{W}_i\right\|^2_F-\frac{1}{2}\left\|\mathbf{Z}_k\right\|^2_F\textcolor{blue}{,~\forall i, k.}
		\end{equation} }Subsequently,~we express the equivalent form of $\text{Rank}(\mathbf{V})\leq 1$  as $\|\mathbf{V}\|_*-	\| \mathbf{V}\|_2 \leq 0$, which is non-convex yet.~Thus, we exploit the iterative MM approach which yields the following convex lower bound: }\begin{algorithm}[t]
	\caption{		{Inner Approximation (IA) Algorithm}}
{\begin{algorithmic}[1]
	\renewcommand{\algorithmicrequire}{\textbf{Input:}}
	\renewcommand{\algorithmicensure}{\textbf{Output:}}
	\REQUIRE $i,t=0$,~$I_\text{max},T_\text{max}$,~$\mathbf{W}^{(t)}_k$,~$\mathbf{W}^{(t)}_\mathrm{E}$,~$\mathbf{{\Theta}}^{(t)}$
	\STATE  $i$ and $t$: Number of iterations
	\STATE  $I_\text{max}$ and $T_\text{max}$: Maximum number of iterations
	\STATE  $\mathbf{\mathbf{W}}^{(t)}_k$: Initial active beamfoming vector in iteration $t$\\
	\STATE  $\mathbf{W}^{(t)}_\mathrm{E}$: Initial energy beam in iteration $t$\\
	\STATE  $\mathbf{\mathbf{\Theta}}^{(t)}$: Initial random phases in iteration $t$\\
	\STATE \textbf{repeat}\\
	\STATE \quad Under given $\mathbf{\Theta}=\mathbf{\Theta}^{(t)}$, $\mathbf{W}^{(t)}_k=\mathbf{W}^{(t)}_k$, and $\mathbf{W}_\mathrm{E}=$\\ \quad $\mathbf{W}^{(t)}_\mathrm{E}$ use \textbf{Algorithm 2} to obtain $\{\rho_{k}^{(t)},\chi^{(t)},\lambda^{(t)}\}$\\
	\STATE \quad Set $\{\rho_{k}^\ast,\chi^\ast,\lambda^*\}=$ $\{\rho_{k}^{(t)},\chi^{(t)},\lambda^{(t)}\}.$
	\STATE \quad\textbf{repeat}\\
	\STATE \quad\quad Calculate $\tilde{F}(\mathbf{W}^{(i)}_i,\mathbf{Z}^{(i)}_k)$, $\tilde{g}(\mathbf{W}^{(i)}_k,\mathbf{Z}^{(i)}_k)$, $\tilde{T}_1(\mathbf{W}_{i})$, \\ \quad\quad$\tilde{T}_2(\mathbf{Z}_{k})$ and $\tilde{F}(\mathbf{W}^{(i)}_{\mathrm{E}},\mathbf{Z}^{(i)}_{k})$ according to (\ref{11}), (\ref{441}),\\ \quad\quad  (\ref{ddd}), (\ref{dddd}), and (\ref{51}),~respectively.
	\STATE \quad\quad  Under given optimal values of problem (P10),~solve\\ \quad\quad problem (P11) and update $\mathbf{V}^{(i+1)}$, $\mathbf{W}^{(i+1)}_k$, and\\ \quad\quad $\mathbf{W}^{(i+1)}_\mathrm{E}$
	\STATE \quad\quad  Decompose $\mathbf{V}^{(i+1)}=\mathbf{v}^{(i+1)}(\mathbf{v}^{(i+1)})^H$ and\\ \quad\quad $\mathbf{W}^{(i+1)}_k$ $=\mathbf{w}^{(i+1)}_k(\mathbf{w}^{(i+1)}_k)^H$
	\STATE \quad\quad Update $\mathbf{\Theta}^{(i+1)}=\text{diag}(\mathbf{v}^{(i+1)})$ 
	\STATE \quad\quad Set ${i}={i}+1$
	\STATE \quad  \textbf{until} ${i}={I}_{\text{max}}$
	\STATE \quad Set $t=t+1$
	\STATE   \textbf{until} $t=T_{{\text{max}}}$.
\end{algorithmic}}
\end{algorithm}
{\begin{align}\label{Taylor}
	\Lambda^{(i)}(\mathbf{V})\triangleq&\|\mathbf{V}\|_*-\|\mathbf{V}^{(i)}\|_2-\text{Tr}\left[\mathbf{v}^{(i)}_{\max}(\mathbf{v}^{(i)}_{\max})^{\text{H}}(\mathbf{V}-\mathbf{V}^{(i)} )\right],
	\end{align}}{where $\mathbf{v}^{(i)}_{\max}$ is the eigenvector corresponding to the maximum eigenvalue of matrix $\mathbf{V}^{(i)}$. Besides, the non-convex constraint (\ref{22c}) can be recast as}
{	\begin{align}\label{18}
\frac{\tilde{\mathbf{X}}^{(i)}_{k,k}}{{{\gamma _k}}} - \sum\limits_{\scriptstyle i = 1\atop\scriptstyle i \ne k}^K {\hat{\mathbf{X}}^{(i)}_{i,k}}  - \sigma _k^2 -\frac{{\delta _k^2}}{{{\rho _k}}}\ge 0\textcolor{blue}{,~\forall k,}
	\end{align}
	where
	\begin{align}
	& {\hat{\mathbf{X}}^{(i)}_{i,k}} \triangleq\frac{1}{2}\left\|\mathbf{W}_i+\mathbf{Z}_k\right\|_F^2-\tilde{T}_1(\mathbf{W}^{(i)}_{i})-\tilde{T}_2(\mathbf{Z}^{(i)}_{k})\textcolor{blue}{,~\forall i, k,}\label{177}\\ 
	&T_1(\mathbf{W}_{i})\triangleq\frac{1}{2}\left\|\mathbf{W}_i\right\|^2_F\geq T_1(\mathbf{W}^{(i)}_{i})\nonumber\\&+\text{Tr}(\nabla_{\mathbf{W}_{i}}^HT_1(\mathbf{W}^{(i)}_{i})(\mathbf{W}_{i}-\mathbf{W}_{i}^{(i)}))\triangleq \tilde{T}_1(\mathbf{W}^{(i)}_{i})\textcolor{blue}{,~\forall i,}\label{ddd}\\
	&T_2(\mathbf{Z}_{k})\triangleq\frac{1}{2}\left\|\mathbf{Z}_k\right\|^2_F\geq T_2(\mathbf{Z}^{(i)}_{k})\nonumber\\&+\text{Tr}(\nabla_{\mathbf{Z}_{k}}^HT_2(\mathbf{Z}^{(i)}_{k})(\mathbf{Z}-\mathbf{Z}^{(i)}_{k}))\triangleq \tilde{T}_2(\mathbf{Z}^{(i)}_{k})\textcolor{blue}{,~\forall  k.}\label{dddd}
	\end{align}}{Similarly, the data rate of each user can be represented as ${R}_k({\mathbf{W}}_i,{\mathbf{Z}}_k)=f({\mathbf{W}}_i,{\mathbf{Z}}_k)-g({\mathbf{W}}_i,{\mathbf{Z}}_k),$~$\textcolor{blue}{\forall i, k.}$  where}
    {\begin{align}
	&{\small f({\mathbf{W}}_k,{\mathbf{Z}}_k)=\text{log}_2\bigg({\sum\limits_{\scriptstyle i= 1}^K {{{ \tilde{\mathbf{X}}_{i,k}}} +\sigma _k^2 +\frac{\delta _k^2}{{\rho _k}}} }\bigg),}\\&
	g({\mathbf{W}}_k,{\mathbf{Z}}_k)=\text{log}_2\bigg({\sum\limits_{\scriptstyle i = 1\atop\scriptstyle i \ne k}^K {{{ {\mathbf{X}}_{i,k} }} + \sigma _k^2 +\frac{\delta _k^2}{{\rho _k}} } }\bigg).
	\end{align}}{Since $g(\mathbf{W}_k,\rho_{k})$ is a differentiable convex function with respect to both $\mathbf{W}_k$ and $\mathbf{Z}_k$,~we have}
{\begin{align}\label{441}
	&g(\mathbf{W}_k,\mathbf{Z}_k)\leq g(\mathbf{W}_k^{(i-1)},\mathbf{Z}_k^{(i-1)})\nonumber\\ &+\text{Tr}\bigg(\big(\nabla_{\mathbf{W}_k}^Hg(\mathbf{W}_k^{(i-1)},\mathbf{Z}_k^{(i-1)})\big)(\mathbf{W}_k-\mathbf{W}_k^{(i-1)})\bigg)\nonumber\\ &+\text{Tr}\bigg(\big(\nabla_{\mathbf{Z}_k}^Hg(\mathbf{W}_k^{(i-1)},\mathbf{Z}_k^{(i-1)})\big)\big(\mathbf{Z}_k-\mathbf{Z}_k^{(i-1)}\big)\bigg)\nonumber\\ &\triangleq\tilde{g}(\mathbf{W}^{(i)}_k,\mathbf{Z}^{(i)}_k)\textcolor{blue}{,~\forall k.}
	\end{align}}{Finally, the following optimization problem can be solved in the $(i+1)$-th iteration as below:}
{	\begin{subequations}
		\begin{align}
		& \text{(P11)}:  \underset{{\mathbf{W}}_k,\mathbf{W}_{\mathrm{E}},\boldsymbol{\Theta}} {\text{minimize}} \:\: \Phi\big(\Lambda(\mathbf{V})\big)\\
		&\text{s.t.}\quad \tilde{R}_{k}({\mathbf{W}}_k,{\mathbf{Z}}_k)-\lambda {{P_{T}}_k}({{{\mathbf{W}}_k,\mathbf{W}_{\mathrm{E}}}}) \geq \chi,~\: \forall k,\\ 
		&\quad\quad \sum\limits_{i = 1}^K \tilde{\mathbf{X}}_{i,k}+\tilde{\mathbf{Y}}_{i,k} \ge \frac{{{P_k}({\text{E}_{\text{min},k}})}}{{1 - {\rho _k}}},~\: \forall k,\\
		&\quad\quad \text{(\ref{power})--(\ref{22g}), (\ref{18})},\quad\text{Diag}(\mathbf{V})=\mathbf{1}_{N+1},\quad \mathbf{V}\succeq\mathbf{0},
		\end{align}
\end{subequations}}	{where}
{\begin{align}
	&\tilde{R}_{k}({\mathbf{W}}_k,{\mathbf{Z}}_k)=f({\mathbf{W}}_k,{\mathbf{Z}}_k)-\tilde{g}({\mathbf{W}}^{(i)}_k,{\mathbf{Z}}^{(i)}_k)\textcolor{blue}{,~\forall k,}\\
	&\tilde{\mathbf{Y}}_{i,k}\triangleq\tilde{F}(\mathbf{W}^{(i)}_{\mathrm{E}},\mathbf{Z}^{(i)}_{k})-\frac{1}{2}\left\|\mathbf{W}_{\mathrm{E}}\right\|^2_F-\frac{1}{2}\left\|\mathbf{Z}_k\right\|^2_F\textcolor{blue}{,~\forall i, k.}\label{51}
	\end{align} }{In addition, $\Phi$ indicates a penalty factor and $\tilde{F}(\mathbf{W}^{(i)}_{\mathrm{E}},\mathbf{Z}^{(i)}_{k})$ can be obtained similar to (\ref{11}). Now, (P11) is a convex optimization problem and optimization tools such as CVX can be exploited to solve it efficiently\cite{CVX}. The Final IA-based algorithm is summarized in Algorithm 3. It is worth mentioning that the objective function in (P11) is non-increasing in each iteration and according to, \cite[Th. 1]{Marks}, the IA-based algorithm is confirmed to converge to a KKT solution of (P1)}. 
\begin{figure}[t]
	\centering
	\includegraphics[width=3.5in]{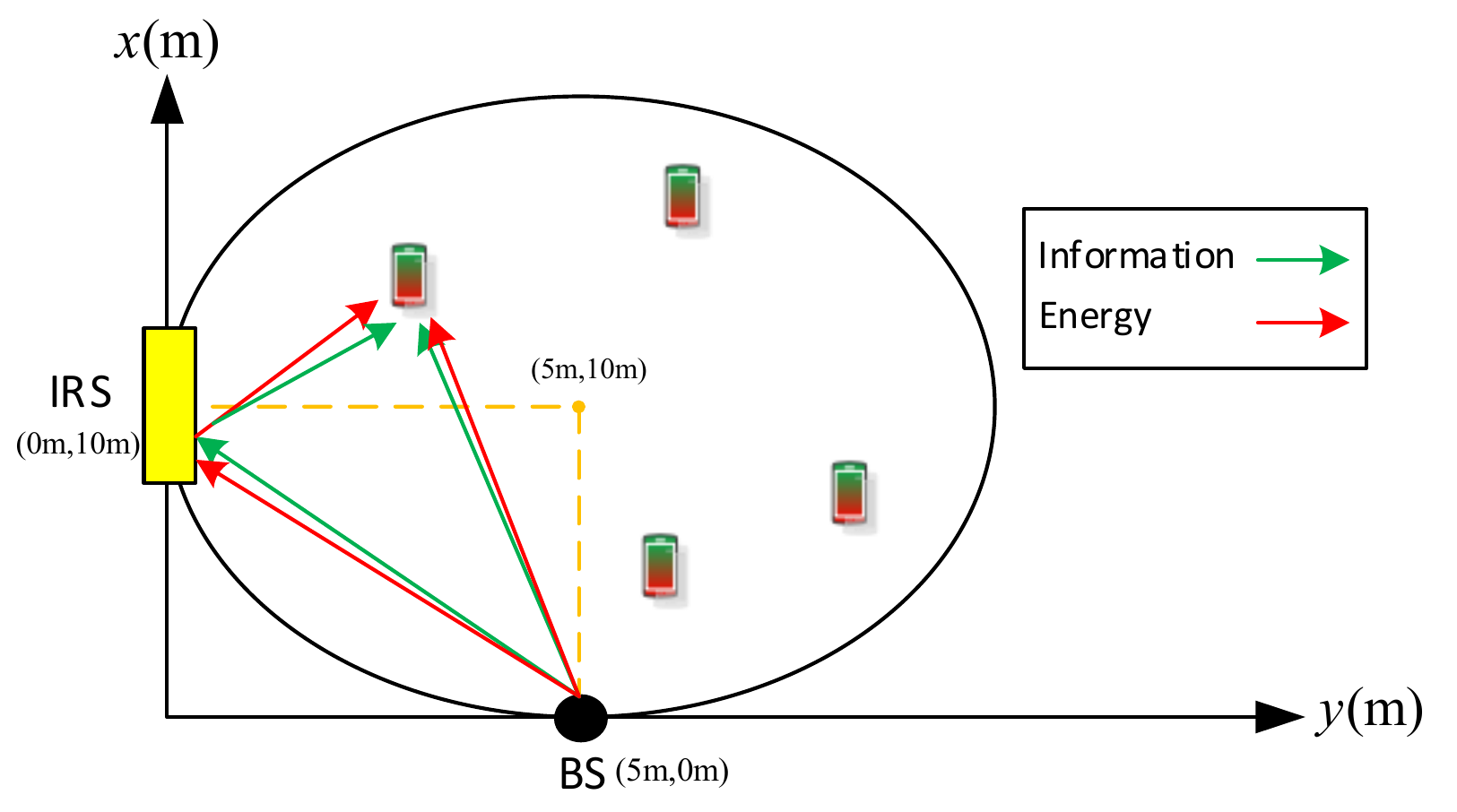}
	\caption{\small The simulated MISO IRS-assisted PS-based SWIPT  communication system.}\label{fig4}
\end{figure} 	
\subsection{Computational Complexity Analysis}

\textcolor{blue}{In this section,~the computational complexity of the proposed algorithms are analyzed.~Note that sub-problems based on the SDP can be solved by the interior-point method.~According to \cite[Th. 3.12]{Bigo},~the order of complexity for a SDP problem with $m$ SDP constraints which includes an $n \times n$ positive \textcolor{blue}{semi-definite} (PSD) matrix is given by $\mathcal{O}\left( \sqrt{n}\log(1/\epsilon)( mn^3+m^2n^2+m^3)\right) $ where $\epsilon\textgreater 0$ is the solution accuracy. For problem (P6),~with $n=M$ and $m=4K+1$,~the approximate computational complexity for solving (P6) can be written as $\mathcal{O}\left( \log(1/\epsilon)((4K+1)(M^{3.5}+4KM^{2.5}))\right)$.~Similarly,~the approximate computational complexity for solving (P8) can be obtained as $\mathcal{O}\left( \log(1/\epsilon)((3K+1)(N^{3.5}+3KN^{2.5}))\right) $. Finally,~the computational complexity of each iteration of Algorithm 3 is asymptotically equal to  $\mathcal{O}\left( \log(1/\epsilon)((M^{3.5}+N^{3.5})K)\right) $. On the other hand, for the second IA-based algorithm,~the asymptotic complexity order of designing joint beamforming and phase shift optimization in each iteration can be expressed as $\mathcal{O}\left( \log(1/\epsilon)(N^{3.5}(M^{3.5}K+M^{2.5}K^2))\right) $.~Given the fact that there is a trade-off between algorithm complexity and system performance, the IA-based algorithm can achieve a better solution at the cost of high complexity order. In contrast, the AO-based algorithm is of much lower complexity.}
	
\section{Simulation Results}
In this section,~numerical results are provided to evaluate the performance of the proposed algorithms.~A two-dimensional (2D) coordinate system is considered in Fig.~\textcolor{blue}{\ref{fig4}},~where a uniform linear array (ULA) and a uniform rectangular array (URA) is adopted at the BS and the IRS,~respectively.~The total number of passive elements at the IRS is assumed to be $N=N_xN_z$,~where $N_x$ and $N_z$ denote the number of phase shifts along the $x$-axis and $z$-axis,~respectively.~The simulation parameters are given in Table II,~unless otherwise is specified.~{Note that the receivers are assumed to be located close to the BS to obtain a higher EH efficiency.} By adopting the Rician fading channel model,~the BS-IRS and IRS-user channels are modeled as
\begin{align}
&\mathbf{H}=\sqrt{\frac{K_r}{1+K_r}}\mathbf{H}^{\text{LOS}}+\sqrt{\frac{1}{1+K_r}}\mathbf{H}^{\text{NLOS}},\label{key11}\\
&\mathbf{h}_{ru,k}=\sqrt{\frac{K_r}{1+K_r}}\mathbf{h}_{ru,k}^{\text{LOS}}+\sqrt{\frac{1}{1+K_r}}\mathbf{h}_{ru,k}^{\text{NLOS}},\label{key10}
\end{align}	
\begin{table}[t]
	\renewcommand{\arraystretch}{1.05}
	\centering
	\caption{\small Simulation Parameters}
	\label{table-notations}
	\begin{tabular}{| c| c| }    
		\hline
		\textbf{Parameters}& \textbf{Values}\\\hline        
		Cell radius & 10~m  \\ \hline
		BS location & (5~m,0~m)  \\ \hline
		IRS location & (0~m,10~m)  \\ \hline   
		{Path loss exponent of NLOS links} & {$\alpha_{\text{NL}}=2.2$}\\ \hline
		{	Path loss exponent of LOS links} & {$\alpha_\text{LOS}=3.6$}\\ \hline
		Rician factor,~$K_r$ & 5~dB \\ \hline
		Received antenna noise power,~$\sigma _k^2$ & $\sigma_k^2=\sigma^2=-70~\text{dBm}$  \\ \hline
		{Carrier frequency,~$f_c$} &{$750\:\text{MHz}$ } \\ \hline 
		Additional noise power at the ID,~$\delta_k^2$ & $\delta_{k}^2=\delta^2=-50~\text{dBm}$    \\ \hline
		Circuit power consumption,~${P_{CR}}_{k}$ & 30~dBm \\ \hline
		Number of antenna at the BS,~$M$ & 5\\ \hline
		Number of reflecting elements,~$N$ & 35\\ \hline
		Number of users,~$K$ & 4\\ \hline
		NL EH model parameters & $a_k=6400$,~$b_k=0.003$\\ \hline
		Penalty factor,~$\mu$ & $5*10^{-5}$\\ \hline
		Maximum harvested
		power, $M_k$ & $0.02$~Watt\\ \hline
		Maximum transmit power, $p_{\max }$ & 30~dBm\\ \hline
		Target SINR, $\gamma_k$ & 10~dB\\ \hline
	\end{tabular}
\end{table}respectively,~where $K_r$ is the Rician factor.~In particular,~$\mathbf{h}_{i,k}^{\text{NLOS}}$ and $\mathbf{H}_{i,k}^{\text{NLOS}}$ denote the non-line-of-sight (NLOS) components which follow Rayleigh fading models.~$\mathbf{H}^{\text{LOS}}$ and $\mathbf{h}_{ru,k}^{\text{LOS}}$ are the line-of-sight (LOS) components which
can be modeled by the ULA.~Besides,~${\small \mathbf{h}_{i,k}^{\text{LOS}}=[1,...,~e^{j(M-1)\theta_k}]^{(t)}}$ where ${\small \theta_k=-\frac{2\pi d\sin(\phi_k)}{\lambda_c}}$,~$\lambda_c$ is the carrier wavelength and $d=\lambda_c/2$ is antenna spacing at the BS.~In particular,~$\mathbf{H}_{i,k}^{\text{LOS}}$ can be defined in the same manner.~{Besides,~it is assumed that the system works on a carrier frequency of $750$ MHz with a wavelength $\lambda_c = 0.4$ m.~The distance-dependent path loss model is given by}
{\begin{align}
L(d) = C_0(\frac{d}{D_0})^{-\alpha},
\end{align}}{where $C_0 =(\frac{\lambda_c}{4\pi})^2$ indicates the path loss at the reference distance $D_0=1$ m,~$d$ is the link distance,~and $\alpha$ denotes the path loss exponent.~The path loss exponents of both the BS-IRS and IRS-user links are assumed to be $\alpha_{\text{NLOS}}=2.2$,~while the AP-user channels is set to $\alpha_\text{LOS}=3.6$.}~For comparison,~we study three benchmark system designs: 
\begin{enumerate}
\item[1)] {IRS-aided SWIPT with AO algorithm where the transmit beamforming vectors and phase shifts are optimized (proposed algorithm) based on the SDR method without penalty method. Note that the SDR scheme has been widely employed in the literature [8], [24], [25].}



\item[2)] IRS-aided SWIPT with random phase shifts at the IRS.

\item[3)] IRS-aided SWIPT with fixed PS ratios.
\end{enumerate}
Note that the transmit beamforming vectors of \textcolor{blue}{2) and 3)} are achieved by employing the SDR technique.

\begin{figure}[t]
	\centering
	\includegraphics[width=12.00cm, height=5.500cm] {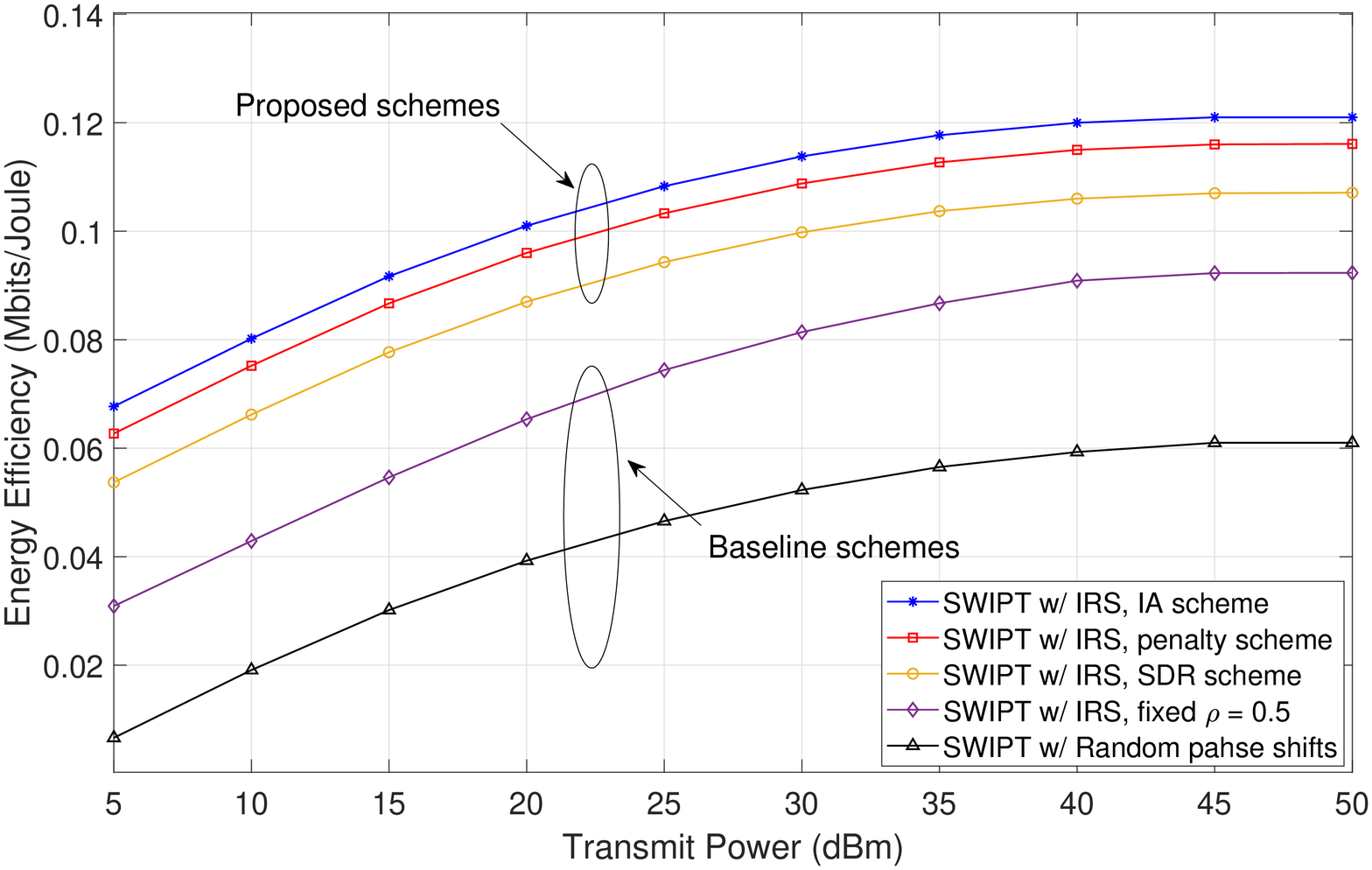}
	\caption{\small EE versus maximum transmit power of BS.}\label{figure_p}
\end{figure}

\subsection{EE versus Maximum Transmit Power}
{Fig.~5 shows the EE versus the maximum allowed transmit power of the BS,~i.e.,~$p_{\max}$ for different schemes.~It can be observed that by increasing maximum transmit power,~EE of all schemes first increases and reaches a maximum value and then remains stable.~This is because for a large value of transmit power,~the interference term increases significantly which degrades the EE of the system.~Also,~for a large value of transmit power,~the exceed transmit power will not be used,~and further increase on the transmit power would degrade the EE of the system.~Besides,~it is also observed that the fixed PS ratios and random phase shifts degrade system performance in the aspect of EE.~An exciting results reveal that EE gain is increased significantly by employing the proposed algorithms as compared to the benchmarks. This is because the two proposed IRS-SWIPT schemes can obtain larger throughput as the IRS operates in FD mode without requiring any SI techniques.~Also,~each user consumes less power since SWIPT helps users harvest energy transmitted by the BS,~which reduces power dissipation.~Other suboptimal schemes require more transmit power in order to satisfy the minimum required energy and rate of each user.~This figure also demonstrates the effectiveness of the proposed algorithm based on IA as compared to SDR schemes as well as penalty scheme in which the phase shift and beamforming are optimized, simultaneously.}	
\begin{figure}[t]
	\centering
	\includegraphics[width=9.400cm, height=5.8000cm] {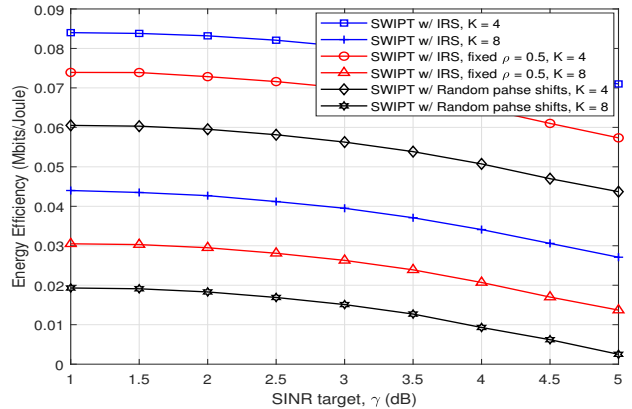}
	\caption{\small EE versus minimum SINR.}\label{figure_gamma}
\end{figure} 
\begin{figure}[t]
\centering
\includegraphics[width=12.000cm, height=5.500cm] {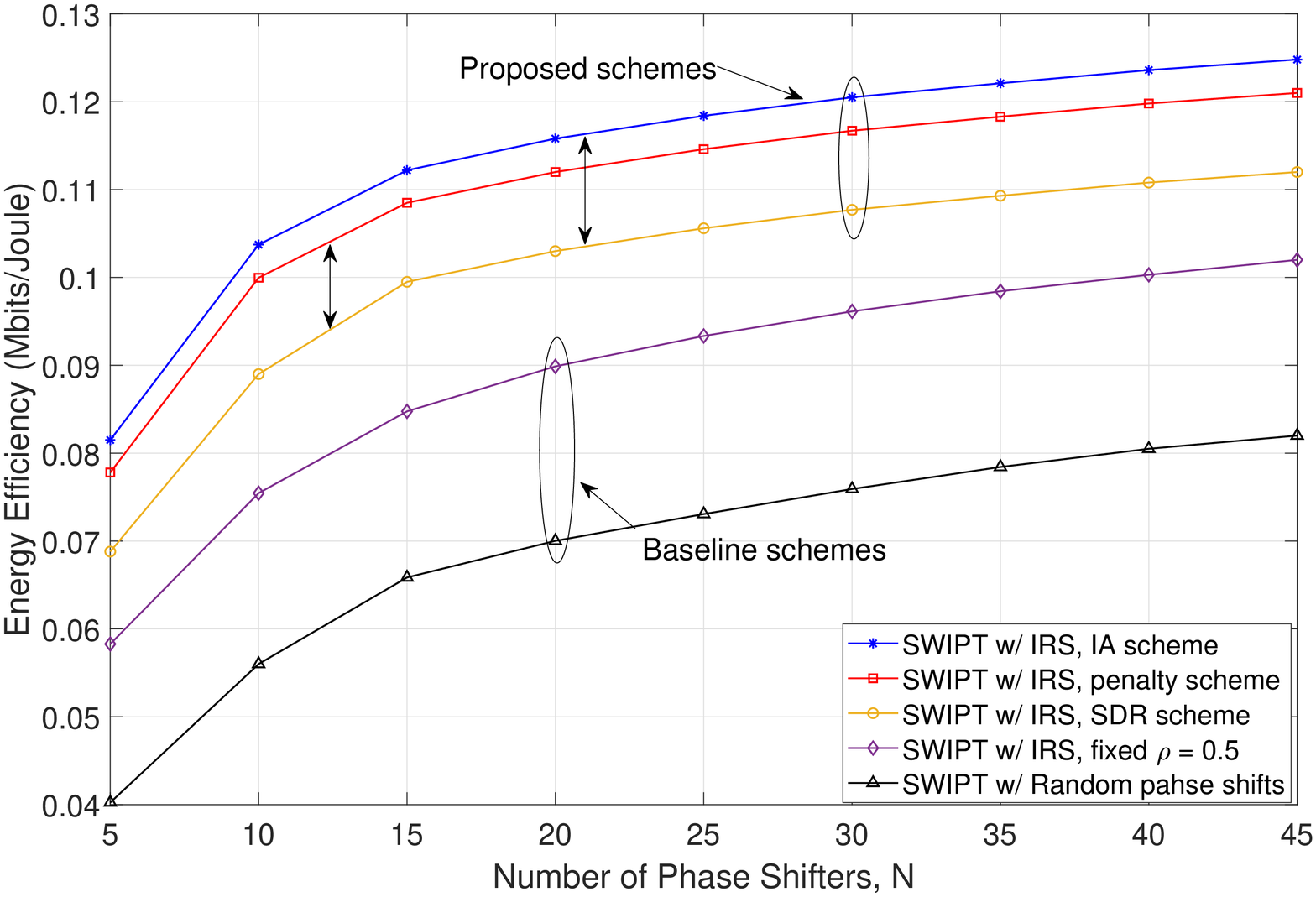}
\caption{\small EE versus number of reflecting elements at the IRS.}\label{figureN}
\end{figure} 
\subsection{EE versus Minimum SINR}
Fig.~6 illustrates the EE versus minimum SINR requirement of each user for different schemes.~It can be observed that by increasing the minimum SINR,~the system EE remains almost the same up to a specific minimum SINR requirement value but drops afterward.~The reason is that for the large value of the minimum SINR,~the BS needs to transmit with high power in order to satisfy the data rate requirement,~which results in magnifying the interference and consequently degrading the system EE.~By deploying the IRS,~the system operates in FD mode and does not need to consume energy to overwhelm the impact of SI.~Moreover,~by considering the IRS-SWIPT,~the proposed algorithm achieves higher max-min EE as compared to fixed PS ratios and random phase shift schemes since the proposed scheme is able to utilize the energy and spectrum effectively. {Indeed,~the proposed AO algorithm can establish a more favorable signal propagation environment by the optimized beamforming at IRS, thus, it considerably outperforms other baseline schemes in terms of EE.}

\subsection{EE versus Number of Reflecting Elements at the IRS}
{Fig.~7 shows max-min EE versus the number of reflecting element at the IRS.~It can be seen that by increasing the number of reflecting elements at IRS,~the max-min EE increases monotonically as well.~This observation can be explained that a large number of $N$ leads to higher SE and much lower aggregated power consumption,~which turns to higher EE of the system.~It is worth noting that for small $N$,~the IRS-SWIPT has a little impact on the performance gain as compared to other schemes.~Indeed,~the small number of $N$ produces a bottleneck,~which indicates that the limited resources of the system restrict the performance gain of the proposed AO algorithm.~However,~since the IRS exploits passive elements,~no radio frequency (RF) chains will be required.~Thus $N$ can be increased with a much lower cost.~As a result,~higher EE gain can be obtained with the aid of IRS than that of the fixed PS ratios and random phase shifts schemes,~mainly when $N$ is adequately large with significantly reduced active RF chains.~This is because a powerful reflective channel link is added by the IRS,~which can help extend the communication range and performing SWIPT at each user.~On the other hand,~the EE achieved by adding IRS to the PS-based SWIPT system increases with $N$,~which highlights the effect of optimizing the phase shifts.~Besides, the performance gain
	obtained by applying the IA method is superior to the penalty
	method as well as the SDR scheme.}

\subsection{EE versus Harvested power}

{Fig.~8 illustrates EE achieved by the proposed algorithm and benchmarks over $\text{E}_\text{min}$ with fixed $K_r=5$ or $K_r=10$. It can be observed that with IRS-aided SWIPT, our proposed solution achieves the maximum EE for all values of the minimum required harvested power, $\text{E}_\text{min}$. Moreover, the EE achieved by the proposed algorithm and benchmarks for $K_r=10$ is notably larger than that by the $K_r=5$ for all values of $\text{E}_\text{min}$. This is because a larger value of $K_r$ indicates a strong LoS path that enables efficient wireless communication. Besides, the IRS with random shifts leads to a small performance improvement compared to the case with fixed $\rho$ since passive beamforming vectors at the IRS are not optimized.}

\begin{figure}[t]\label{figureH}
	\centering
	\includegraphics[width=9.700cm, height=6.500cm] {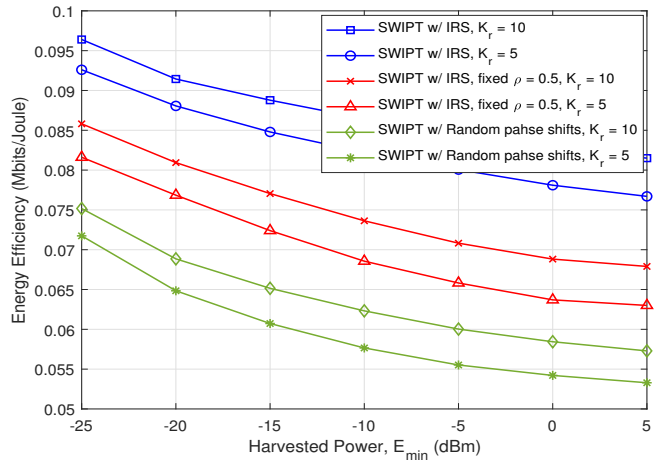}
	\caption{\small EE versus harvested power.}
\end{figure}
%

%
\begin{figure}[t]
	\centering
	\includegraphics[width=9.500cm, height=6.500cm] {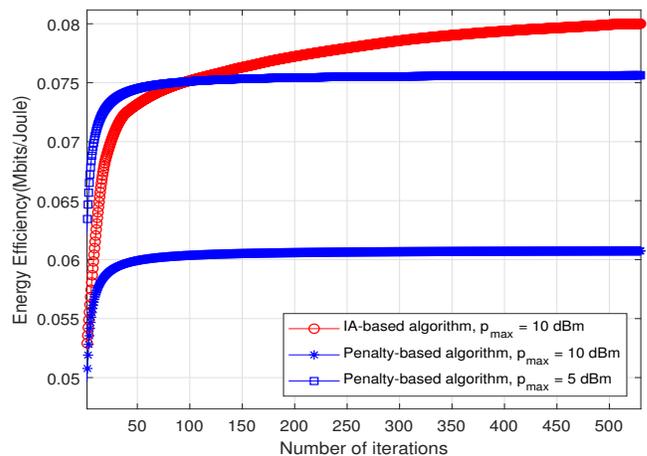}
	\caption{\small Convergence behavior of the proposed algorithm for solving problem (P1).}
\label{figureI}\end{figure} 
\subsection{Convergence of Iterative Algorithm}
{	Fig.~10 depicts the system EE versus the number of iterations with different values of maximum transmitted powers for the penalty- and IA-based algorithms. As can be observed both algorithms ensure a monotonic convergence while the speed of convergence differs from one case to another. \textcolor{blue}{It can be perceived that the EE of the system based on the 
		IA-based algorithm requires more iterations to converge.} In all cases, our proposed algorithm converges to a stationary point only after a number of iterations which shows the effectiveness of the proposed algorithm.}
\section{Conclusion}
This paper studied the joint transmit and reflect beamforming as well as receive PS ratios design for a multi-user MISO IRS-aided SWIPT with an additional energy beam. In particular,~the max-min EE was maximized subject to the SINR and EH constraints.~To tackle it,~we proposed an efficient algorithm to solve the non-convex problem by applying the AO algorithm,~DC programming,~and SDR as well as MM techniques,~which then prove its tightness.~{Beside, a new algorithm based on jointly beamforming vectors and phase shifts optimization was provided known as IA method to strike a balance between convergence rate and performance gain}~Simulation results revealed useful insights on the deployment of IRS-aided SWIPT,~which improved the system performance in EE as compared to other schemes. {Another interesting direction for future work is an extension of our work to a more general case by taking into account the imperfect CSI.}	

\appendix

{	\subsection{Proof of Proposition 1}}
	{The approximation in (22) produces a tight lower bound of ${R}_{k}({\mathbf{W}}_k,\rho_{k})$. This is because ${g}(\mathbf{W}_k,\rho_k)$ is a concave function. The gradient
	of ${g}(\mathbf{W}_k,\rho_k)$ is a supper-gradient [34] given by
	\begin{equation}
	{g}(\mathcal{H})\leq\tilde{g}(\mathcal{H}),
	\end{equation}
	where $\mathcal{H}=\{\mathbf{W}_k,\rho_{k}\}$ indicates the set of feasible solutions at iteration $t$.~It can be inferred that $f(\mathcal{H})-{g}(\mathcal{H}) \geq f(\mathcal{H})-\tilde{g}(\mathcal{H})$.~Also,~the equality holds when $\mathbf{W}_k=\mathbf{W}_k^{(t-1)}$ and $\rho_{k}=\rho_{k}^{(t-1)}$,~which confirms the tightness of the lower bound.~First,~we rewrite the objective function of (P4) at iteration $t$ as ${f(\mathcal{H}^{(t)})}-{\tilde{g}(\mathcal{H}^{(t)})}$.~Consequently, we have the following relations:
	{\small \begin{align}
	&f(\mathcal{H}^{(t+1)})-g(\mathcal{H}^{(t+1)})\geq f(\mathcal{H}^{(t+1)})-g(\mathcal{H}^{(t)})\\\nonumber
	&-\text{Tr}\big(\nabla_{\mathbf{W}_k}^Hg(\mathcal{H}^{(t)})(\mathbf{W}_k^{(t+1)}-\mathbf{W}_k^{(t)})\big)-\partial _{\rho_k}g(\mathcal{H}^{(t)})\big(\rho_k^{(t+1)}-\rho_k^{(t)}\big)\nonumber\\
	&=\max_{\mathbf{W}_k,\rho_{k}}f(\mathcal{H})-g(\mathcal{H}^{(t)})-\text{Tr}\big(\nabla_{\mathbf{W}_k}^Hg(\mathcal{H}^{(t)})(\mathbf{W}_k-\mathbf{W}_k^{(t)})\big)\nonumber\\\nonumber
	&-\partial _{\rho_k}g(\mathcal{H}^{(t)})\big(\rho_k-\rho_k^{(t)}\big)\geq f(\mathcal{H}^{(t)})-g(\mathcal{H}^{(t)})\\ \nonumber &-\text{Tr}\big(\nabla_{\mathbf{W}_k}^Hg(\mathcal{H}^{(t)})(\mathbf{W}_k^{(t)}-\mathbf{W}_k^{(t)})\big)-\partial _{\rho_k}g(\mathcal{H}^{(t)})\big(\rho_k^{(t)}-\rho_k^{(t)}\big)\\ \nonumber&=f(\mathcal{H}^{(t)})-g(\mathcal{H}^{(t)}).
	\end{align}}
	Therefore, by solving the convex lower bound in (P4), the iterative-based SCA algorithm creates a sequence of feasible solutions, i.e., $\mathbf{W}_k^{(t+1)}$ and $\rho_{k}^{(t+1)}$. In other words, the solution is monotonically increasing over each iteration.}

\subsection{Proof of Theorem 1}
This theorem can be proved by verifying the sufficiency and necessity criteria.~First,~we prove the sufficiency criterion.~We assume that  $\mathcal{H}^\ast$ is the optimal solution for (\ref{13}) and for any feasible solution,~we have
\begin{align}
&\underset{k}{\text{min}}\:\: \left\{{R_{k}(\mathcal{H})}-\lambda^{\text{opt}}{{{{P_{T}}_k}(\mathcal{F})}}\right\}\leq 0,\label{38}\\
&\underset{k}{\text{min}}\:\: \left\{{R_{k}(\mathcal{H}^\ast)}-\lambda^{\text{opt}}{{{{P_{T}}_k}(\mathcal{F}^{\text{opt}})}}\right\}= 0,\label{39}
\end{align}
where $({\mathbf{W}}_k,\mathbf{W}_{\mathrm{E}}) \in\mathcal{F}$ and $(\rho_{k},{\mathbf{W}}_k) \in\mathcal{H}$ are the sets of feasible solutions. Form the equations (\ref{38}) and (\ref{39}),~we can conclude that
\begin{equation}
\underset{k}{\text{min}}\:\: \frac{{R_{k}(\mathcal{H})}}{{{{P_{T}}_k}(\mathcal{F})}}\leq \lambda^{\text{opt}},~ \quad \underset{k}{\text{min}}\:\: \frac{{R_{k}(\mathcal{H}^*)}}{{{{P_{T}}_k}(\mathcal{F}^{\text{opt}})}}= \lambda^{\text{opt}}.
\end{equation}
Thus,~$\mathcal{H}^*$ is also the optimal solution of (P1),~which completes the sufficiency criterion proof.~Now,~we focus on proving the necessity criterion.~For any feasible solution,~we have
\begin{equation}\label{41}
\underset{k}{\text{min}}\:\: \frac{{R_{k}(\mathcal{H})}}{{{{P_{T}}_k}(\mathcal{F})}}\leq \lambda^{\text{opt}}, \:\:\:\:
\underset{k}{\text{min}}\:\: \frac{{R_{k}(\mathcal{H}^{\text{opt}})}}{{{{P_{T}}_k}(\mathcal{F}^{\text{opt}})}}= \lambda^{{\text{opt}}}.
\end{equation}
By rearranging (\ref{41}),~we obtain
\begin{align}
&\underset{k}{\text{min}}\:\: \left\{{R_{k}(\mathcal{H})}-\lambda^{\text{opt}}{{{{P_{T}}_k}(\mathcal{F})}}\right\}\leq 0,\label{42}\\
&\underset{k}{\text{min}}\:\: \left\{{R_{k}(\mathcal{H}^{\text{opt}})}-\lambda^{\text{opt}}{{{{P_{T}}_k}(\mathcal{F}^{\text{opt}})}}\right\}= 0.\label{43}
\end{align}
Accordingly,~$\mathcal{H}^{\text{opt}}$ is also the optimal solution of (\ref{13}),~which completes the necessity criterion proof.
\subsection{Proof of Proposition 2}
It is worth mentioning that (P6) is a convex optimization problem and also satisfies the Slater’s condition.~In this case,~the duality gap between the primal and dual problem is zero.~Before proceeding to express the Lagrangian function,~we first rewrite (P6) as follows:
\begin{subequations}
	\label{Primal}
	\begin{align}
	&\quad\quad \underset{\rho_{k},{\mathbf{W}}_k,\mathbf{W}_{\mathrm{E}},\chi} {\text{maximize}} \: \chi\\
	&\text {s.t.} \quad \text{log}_2(\varphi_k+\sigma _k^2 +\frac{\delta _k^2}{{\rho _k}})-\tilde{g}(\mathbf{W}_k,\rho_k) \geq \lambda \text{Tr}({\mathbf{W}}_k)\nonumber\\
	&\quad\quad+\lambda \text{Tr}(\mathbf{W}_\mathrm{E})+\lambda {P_{CR}}_{k}+\chi,\: \forall k,\label{34b}\\
	&\quad\quad    {{{ {\text{Tr}({{\mathbf{H}}_k}{{\mathbf{W}}_k})}}}}+{\sum\limits_{ i \ne k} {{{ {\text{Tr}({{\mathbf{H}}_k}{{\mathbf{W}}_i})}}}} }\geq \varphi_k,\: \forall k,\\
	&\quad\quad \text{(\ref{22c})-(\ref{22g})},
	\end{align}
\end{subequations}
where constraint (\ref{34b}) is written in its epigraph form.~The mentioned problem above is jointly convex with respect to the optimization variables and satisfies the Slater’s condition.~Hence,~the Lagrangian function\footnote{In order to facilitate the notations,~without loss of generality,~we consider $\nabla_{\mathbf{W}_k}g(\mathbf{W}_k^{(t-1)},\rho_k^{(t-1)})$ instead of $\nabla_{\mathbf{W}_k}g$ in this proof.} can be defined as (\ref{Lagrange}) where $\{\Gamma_k,\psi_k,\beta_k,\alpha,\mu_k \}$ denote the Lagrangian multipliers associated with the constraints of (P6).~Consequently,~the dual function of the problem (\ref{Primal}) is given by
\begin{figure*}
	\begin{align}\label{Lagrange}
	\displaystyle  
	\mathcal{L}=&\chi+\sum\limits_{k = 1}^K \Gamma_k\bigg(\log_{2}(\varphi_k+\sigma _k^2 +\frac{\delta _k^2}{{\rho _k}})-\text{Tr}(\mathbf{W}_k\nabla_{\mathbf{W}_k}^Hg)-\lambda \text{Tr}(\mathbf{W}_k)-\lambda \text{Tr}(\mathbf{W}_\mathrm{E})-\lambda {P_{CR}}_{k}-\chi\bigg)\nonumber\\&+\sum\limits_{k = 1}^K \psi_k\bigg(\frac{{\text{Tr}({{\mathbf{H}}_k}{{\mathbf{W}}_k})}}{{{\gamma _k}}} - \sum\limits_{\scriptstyle i = 1\atop\scriptstyle i \ne k}^K {\text{Tr}({{\mathbf{H}}_k}{{\mathbf{W}}_i})} - \sigma _k^2-\frac{{\delta _k^2}}{{{\rho _k}}} \bigg)
	+\sum\limits_{k = 1}^K \beta_k \bigg(\sum\limits_{i = 1}^K {{{ \text{Tr}({{\mathbf{H}}_k}{{\mathbf{W}}_i}) }}}+\text{Tr}({{\mathbf{H}}_k}\mathbf{W}_\mathrm{E})- \frac{{{P_k}({\text{E}_{\text{min},k}})}}{{1 - {\rho _k}}} \bigg)
	\nonumber\\&-\alpha \bigg(\sum\limits_{k = 1}^K {\text{Tr}({{\mathbf{W}}_k})}+{\text{Tr}({{\mathbf{W}}_E})}  - p_\text{max} \bigg) +\sum\limits_{k = 1}^K \mu_k \bigg(     {{{ {\text{Tr}({{\mathbf{H}}_k}{{\mathbf{W}}_k})}}}} +{\sum\limits_{\scriptstyle i= 1\atop\scriptstyle i \ne k}^K {{{ {\text{Tr}({{\mathbf{H}}_k}{{\mathbf{W}}_i})}}}} }-\varphi_k \bigg).
	\end{align}
	\hrule
\end{figure*}
\begin{subequations}
	\label{daul}
	\begin{align}
	& \underset{\rho_{k},{\mathbf{W}}_k,\mathbf{W}_{\mathrm{E}},\chi} {\text{maximize}} \:\: \mathcal{L}\big(\rho_{k},{\mathbf{W}}_k,\mathbf{W}_{\mathrm{E}},\chi,\Gamma_k,\psi_k,\beta_k,\alpha,\mu_k \big)\\
	&\text {s.t.} \quad {\mathbf{W}}_k\succeq \mathbf{0},\: \mathbf{W}_{\mathrm{E}}\succeq\mathbf{0},\:\: \forall k,\\
	&\quad\quad 0<\rho_{k}<1,\:\: \forall k.
	\end{align}
\end{subequations}    
Accordingly,~problem (\ref{daul}) can be explicitly expressed by (\ref{47}) where
\begin{figure*}
	\centering
	\begin{minipage}{1\textwidth}
		\begin{align}\label{47}
		\displaystyle 
		\mathop {\text{minimize} }\limits_{\scriptstyle {\mathbf{W}}_k\succeq \mathbf{0},\mathbf{W}_{\mathrm{E}}\succeq\mathbf{0} \atop\scriptstyle 0<\rho_{k}<1,\chi} \: &\bigg\{ \sum\limits_{\scriptstyle k= 1\atop\scriptstyle}^K \text{Tr}(\mathbf{A}_k\mathbf{W}_k)+\text{Tr}(\mathbf{C}\mathbf{W}_\mathrm{E})+ \sum\limits_{\scriptstyle k= 1\atop\scriptstyle}^K \bigg(\psi_k(\frac{\delta _k^2}{{\rho _k}})-\Gamma_k\log_{2}(\varphi_k+\sigma _k^2 +\frac{\delta _k^2}{{\rho _k}}) + \beta_k\bigg(\frac{{{P_k}({\text{E}_{\text{min},k}})}}{{1 - {\rho _k}}}\bigg)\bigg)\nonumber\\&+\sum_{k=1}^{K}\psi_k\sigma_{k}^2+\sum_{k=1}^{K} \mu_k\varphi_k+\sum_{k=1}^{K}\Gamma_k\bigg(\lambda {P_{CR}}_{k}+\chi\bigg)-\alpha p_\text{max}-\chi\bigg\}.
		\end{align}
		\medskip
		\hrule
	\end{minipage}
\end{figure*}
\begin{align}
&\mathbf{A}_k=\text{Tr}(\Gamma_k\nabla_{\mathbf{W}_k}^Hg)+\Gamma_k\lambda \mathbf{I}_M-\frac{\psi_k}{\gamma_{k}}\mathbf{H}_k+\sum_{i}\psi_i\mathbf{H}_i\nonumber\\&-\psi_k\mathbf{H}_k-\sum_{i}\beta_i\mathbf{H}_i+\alpha \mathbf{I}_M-\sum_{i} \mu_i\mathbf{H}_i,\label{48}\\
&\mathbf{C}=(\Gamma_k \lambda +\alpha )\mathbf{I}_M-\sum_{k}\beta_k\mathbf{H}_k.\label{49}
\end{align}
Let $\{\Gamma_k^\ast,\psi_k^\ast,\beta_k^\ast,\alpha^\ast,\mu_k^\ast \}$ indicate the optimal dual solution set to the problem (\ref{daul}).~Therefore,~we have	
\begin{align}
&\mathbf{A}_k^\ast=(\Gamma_k^\ast\lambda +\alpha^\ast)\mathbf{I}_M+\text{Tr}(\Gamma_k^\ast\nabla_{\mathbf{W}_k}^Hg)+\sum_{i}(\psi_i^\ast-\beta_i^\ast\nonumber\\&-\mu_i^\ast)\mathbf{H}_i-(\frac{\psi_k^\ast}{\gamma_{k}}+\psi_k^\ast)\mathbf{H}_k,\label{AA}\\
&\mathbf{C}^*=(\Gamma_k^\ast \lambda +\alpha^\ast )\mathbf{I}_M-\sum_{k}\beta_k^\ast\mathbf{H}_k.\label{BB}
\end{align}
Then by considering (\ref{AA}) and (\ref{BB}),~it can be observed that under given any $k$,~$\mathbf{W}_k^\ast$ and $\mathbf{W}_\mathrm{E}^\ast$ are the solutions to the following problems:
\begin{subequations}
	\begin{align}
	&\underset{{\mathbf{W}}_k\succeq \mathbf{0}} {\text{minimize}} \:\: \text{Tr}(\mathbf{A}_k^\ast\mathbf{W}_k),\\    &\underset{{\mathbf{W}}_E\succeq \mathbf{0}} {\text{minimize}} \:\: \text{Tr}(\mathbf{C}^\ast\mathbf{W}_\mathrm{E}).
	\end{align}
\end{subequations}
Note that,~we have $\mathbf{A}_k^\ast\succeq\mathbf{0}$,~$\forall k$,~and $\mathbf{C}^\ast\succeq\mathbf{0}$ to ensure bounded dual optimal solutions.~By utilizing (\ref{AA}) together with $\mathbf{A}_k^\ast\succeq\mathbf{0}$ and $\mathbf{C}^\ast\succeq\mathbf{0}$,~we obtain
\begin{align}
&\text{Tr}(\mathbf{A}_k^\ast\mathbf{W}_k)=0,\:\: \forall k,\label{53}\\
&\text{Tr}(\mathbf{C}^\ast\mathbf{W}_\mathrm{E})=0\label{54},
\end{align}
which are the complementary slackness conditions.~Furthermore,~the optimal PS solution $\rho_k$,~$\forall k$,~can be obtained from (\ref{Primal}),~which is the solution to the following problem:
\begin{equation}\label{55}
\underset{0<\rho_{k}<1} {\text{max}} \:  \sum\limits_{\scriptstyle k= 1\atop\scriptstyle}^K \Gamma_k\log_{2}(\varphi_k+\sigma _k^2 +\frac{\delta _k^2}{{\rho _k}})-\psi_k(\frac{\delta _k^2}{{\rho _k}})-\beta_k\bigg(\frac{{{P_k}({\text{E}_{\text{min},k}})}}{{1 - {\rho _k}}}\bigg).
\end{equation}
Please note that in the problem (\ref{55}),~if $\psi_k^\ast=0$ and $\beta_k^\ast\geq0$,~then the optimal PS tends to infinity i.e.,~$\rho_{k}^\ast \to \infty$.~In the same manner,~if $\psi_k^\ast\geq0$ and $\beta_k^\ast=0$,~the optimal PS tends to one i.e.,~$\rho_{k}^\ast\to1$.~As a result,~neither of the two cases is true since $0<\rho_{k}^\ast<1$,~$\forall k$,~must hold.~Moreover,~$\psi_k^\ast=0$ and $\beta_k^\ast=0$ can not be valid for any $k$ which will be shown by contradiction in the following.~In this case,~assume that there is some $k$’s so that $\psi_k^\ast=\beta_k^\ast=0$.~Let us define the following set:
\begin{equation}
\Omega\triangleq\{k|\psi_k^\ast=\beta_k^\ast=0,~\:1\leq k \leq K\},
\end{equation}
where $\Omega\neq0$.~Next,~by defining $\mathbf{B}^\ast=(\Gamma_k^\ast\lambda +\alpha^\ast)\mathbf{I}_M+\text{Tr}(\Gamma_k^\ast\nabla_{\mathbf{W}_k}^Hg)+\sum\limits_{i\notin \Omega}(\psi_i^\ast-\beta_i^\ast-\mu_i^\ast)\mathbf{H}_i$,~relation (\ref{AA}) can be written as
\begin{equation}\label{57}
\mathbf{A}^\ast_{k}= \left\{ \begin{array}{l}\mathbf{{B}^\ast},{\rm{      \quad\quad\quad\quad\quad \quad\quad \:\:\:\:\: if \: k}} \in \Omega ,\\{\mathbf{B}^\ast} -(\frac{\psi_k^\ast}{\gamma_{k}}+\psi_k^\ast)\mathbf{H}_k,{\rm{    \quad\text{otherwise}}}{\rm{.}}\end{array} \right.
\end{equation}
Considering the fact that $\mathbf{A}_k\succeq\mathbf{0}$ and $(\frac{\psi_k^\ast}{\gamma_{k}}+\psi_k^\ast)\mathbf{H}_k\succeq\mathbf{0}$,~$\forall k$, it can be observed that $\mathbf{B}^\ast\succeq\mathbf{0}$.~In the subsequent,~we prove that $\mathbf{B}^\ast$ is positive definite by contradiction.~Suppose $\mathbf{B}^\ast$ is a positive \textcolor{blue}{semi-definite} matrix which the minimum eigenvalue of $\mathbf{B}^\ast$ is assumed to be zero.~Hence,~there is at least one $\mathbf{x}\neq\mathbf{0}$ such that $\mathbf{x}^H\mathbf{B}^\ast \mathbf{x}=0$.~As a result,~we have
\begin{equation}
\mathbf{x}^H\mathbf{A}_k^\ast \mathbf{x}=(\frac{\psi_k^\ast}{\gamma_{k}}+\psi_k^\ast)\mathbf{x}^H\mathbf{H}_k\mathbf{x}\leq0,\:\: k\notin\Omega.
\end{equation}
Since $\psi_k^\ast>0$,~it leads to ${\mathbf{x}}^H\mathbf{H}_k\mathbf{x}=0$,~$k\notin\Omega$.~Therefore,~we obtain
\begin{align}\label{59}
&\mathbf{x}^H\mathbf{B}^\ast \mathbf{x}=\mathbf{x}^H\bigg((\Gamma_k^\ast\lambda +\alpha^\ast)\mathbf{I}_M+\text{Tr}(\Gamma_k^\ast\nabla_{\mathbf{W}_k}^Hg)\nonumber\\&+\sum_{i\notin\Omega}(\psi_i^\ast-\beta_i^\ast-\mu_i^\ast)\mathbf{H}_i\bigg)\mathbf{x}=\mathbf{x}^H(\Gamma_k^\ast\lambda +\alpha^\ast)\mathbf{x}>0,\:\: k\notin\Omega,
\end{align}
where $\alpha^\ast>0$ must hold for $p_\text{max}>0$.~It can be observed that,~the relation (\ref{59}) contradicts $\mathbf{x}^H\mathbf{B}^\ast \mathbf{x}=\mathbf{0}$,~which follows that $\mathbf{B}^\ast\succ\mathbf{0}$ and $\text{Rank}(\mathbf{B}^\ast)=M$.~Based on (\ref{57}),~it follows that if $k\in\Omega$,~then $\text{Rank}(\mathbf{A}_k)=M$ can not be true since $\mathbf{W}_k=\mathbf{0}$ is not an optimal solution.~Hence,~we find that $\Omega=0$ which means that $\psi_k^\ast=\beta_k^\ast=0$,~$\forall k$,~is not a valid dual solution.~In summary,~it leads to the conclusion that $\psi_k^\ast>0$ and $\beta_k^\ast>0$ must satisfy for any $k$.~To this point,~it has been shown that $\text{Rank}(\mathbf{B}^\ast)=M$ such that $\mathbf{A}_k^\ast={\mathbf{B}^\ast} -(\frac{\psi_k^\ast}{\gamma_{k}}+\psi_k^\ast)\mathbf{H}_k$ results to $\text{Rank}(\mathbf{A}_k^\ast)\geq M-1$,~$\forall k$.~If we consider $\mathbf{A}_k^\ast$ to be full rank,~then according to the (\ref{53}),~we obtain $\mathbf{W}_k^\ast=0$,~$\forall k$, which is not an optimal solution.~Therefore,~we have $\text{Rank}(\mathbf{A}_k^\ast)=M-1$,~$\forall k$, which leads to $\text{Rank}(\mathbf{W}_k^\ast)=1$,~$\forall k$. 

Now we focus on proving $\text{Rank}(\mathbf{W}_\mathrm{E}^\ast)\leq 1$.
Since $\alpha^\ast>0$ and all the channel vectors are independently distributed,~$(\Gamma_k^\ast \lambda +\alpha^\ast )\mathbf{I}_M$ spans the whole signal space such that $\text{Rank}\big((\Gamma_k^\ast \lambda +\alpha^\ast )\mathbf{I}_M\big)=M$.~Moreover,~according to the rank property of the matrices,~we have
\begin{align}\label{60}
&\text{Rank}(\mathbf{C})+\text{Rank}\bigg(\sum_{k}\beta_k^\ast\mathbf{H}_k\bigg)\geq\text{Rank}\bigg((\Gamma_k^\ast \lambda +\alpha^\ast )\mathbf{I}_M\bigg)\nonumber\\
&\Rightarrow\text{Rank}(\mathbf{C})\geq M-1.
\end{align}
It can be observed that,~there exists two cases according to (\ref{60}),~i.e.,~$\text{Rank}(\mathbf{C}^\ast)=M$ and $\text{Rank}(\mathbf{C}^\ast)=M-1$.~For the first case,~it follows that $\mathbf{W}^\ast_\mathrm{E}=\mathbf{0}$ according to the (\ref{54}).~For the second case,~$\text{Rank}(\mathbf{C}^\ast)=M-1$ results in $\text{Rank}(\mathbf{W}^\ast_\mathrm{E})\leq1$ which means that $\mathbf{W}^\ast_\mathrm{E}$ lies in the null space of $\mathbf{C}^\ast$.~Thus,~$\text{Rank}(\mathbf{W}^\ast_\mathrm{E})\leq1$ needs to be satisfied.~On other words,~one energy beam is required at most to achieve the system design goal.

\subsection{{Proof of Proposition 3}}
{In this section,~we aim at proofing the convergence of proposed Algorithm 3.~Let us consider $\{\mathbf{\Theta}^{(j+1)},\mathbf{W}_k^{(j)},\mathbf{W}_\mathrm{E}^{(j)},\rho_k^{(j)}\}$ as the feasible solution set to (P7),~then it is a feasible solution to (P6) as well.~Accordingly,~$\{\mathbf{\Theta}^{(j)},\mathbf{W}_k^{(j)},\mathbf{W}_\mathrm{E}^{(j)},\rho_k^{(j)}\}$  and $\{\mathbf{\Theta}^{(j+1)},\mathbf{W}_k^{(j+1)},\mathbf{W}_\mathrm{E}^{(j+1)},\rho_k^{(j+1)}\}$ are feasible to (P6) in the $j$-th and $(j+1)$-th iterations,~respectively.~Besides,~we express the objective value of (P6) as $f(\mathbf{\Theta},\mathbf{W}_k,\mathbf{W}_\mathrm{E},\rho_k)$.~Consequently,~we obtain
	{\small \begin{equation}
	f(\mathbf{\Theta}^{(j+1)},\mathbf{W}_k^{(j+1)},\mathbf{W}_\mathrm{E}^{(j+1)},\rho_k^{(j+1)})\geq f(\mathbf{\Theta}^{(j+1)},\mathbf{W}_k^{(j)},\mathbf{W}_\mathrm{E}^{(j)},\rho_k^{(j)}).
	\end{equation}}
	Thus, for given phase shifts,~$\mathbf{\Theta}^{(j+1)}$,~the set of solutions $\{\mathbf{W}_k^{(j+1)},\mathbf{W}_\mathrm{E}^{(j+1)},\rho_k^{(j+1)}\}$ are suboptimal.~Furthermore,~we have 
	\begin{equation}
	f(\mathbf{\Theta}^{(j+1)},\mathbf{W}_k^{(j)},\mathbf{W}_\mathrm{E}^{(j)},\rho_k^{(j)})\geq f(\mathbf{\Theta}^{(j)},\mathbf{W}_k^{(j)},\mathbf{W}_\mathrm{E}^{(j)},\rho_k^{(j)}).
	\end{equation}
	Considering that the objective function value dose not depend on $\mathbf{\Theta}$.~As a result,~it follows that 
	\begin{equation}
	f(\mathbf{\Theta}^{(j+1)},\mathbf{W}_k^{(j+1)},\mathbf{W}_\mathrm{E}^{(j+1)},\rho_k^{(j+1)})\geq f(\mathbf{\Theta}^{j},\mathbf{W}_k^{j},\mathbf{W}_\mathrm{E}^{j},\rho_k^{j}),
	\end{equation}
	It should be noted that since the initial point of each step is actually the starting point of the previous round,~better EE is guaranteed.~Since these steps proceed such that at each round,~a better EE is achieved,~and the initial point of each step corresponds to the latest achievable EE; hence,~the EE would be increased.
	It can be perceived that at each iteration the parameters would be updated based on the results from the previous iteration in which the value of the objective function would be improved or at least would be remained unchanged with respect to the previous iteration,~which completes the proof.}


\begin{thebibliography}{99}
	\bibitem{1}Q. Wu,~G. Y. Li,~W. Chen,~D. W. K. Ng,~and R. Schober,~“An overview of sustainable green 5G networks,” \textit{IEEE Wireless Commun.},~vol. 24,~no. 4,~pp. 72–80,~Aug. 2017.
	
	
	\bibitem{2}S. Buzzi,~C. I,~T. E. Klein,~H. V. Poor,~C. Yang,~and A. Zappone,~“A survey of energy-efficient techniques for 5G networks and challenges ahead,” {\textit{IEEE J. Sel. Areas Commun.}},~vol. 34,~no. 4,~pp. 697–709,~Apr. 2016.
	
	\bibitem{3}S. Zhang,~Q. Wu,~S. Xu,~and G. Y. Li,~“Fundamental green trade-offs: Progresses,~challenges,~and impacts on 5G networks,” \textit{IEEE Commun. Surveys Tuts.},~vol. 19,~no. 1,~pp. 33–56,~First Quarter 2017.
	
	
	
	
	\bibitem{77}Q. Wu and R. Zhang,~“Towards smart and reconfigurable environment: Intelligent reflecting surface aided wireless networks,” \textit{IEEE Commun. Mag.},~vol. 58,~no. 1,~pp. 106-112,~Jan. 2020.
	
	
	\bibitem{Wu44}S. Abeywickrama,~R. Zhang,~Q. Wu,~and C. Yuen,~“Intelligent reflecting surface: Practical phase shift model and beamforming optimization,”\textit{ IEEE Trans. Commun.}, vol. 68, no. 9, pp. 5849-5863, Sep. 2020.
	
	
	\bibitem{7}E. Basar,~M. Di Renzo,~J. De Rosny,~M. Debbah,~M. Alouini,~and R. Zhang,~“Wireless communications through reconfigurable intelligent surfaces,” \textit{IEEE Access},~vol. 7,~pp. 116 753–116 773,~Aug. 2019.
	
	
	
	
	\bibitem{Discrete} Q. Wu and R. Zhang, “Beamforming optimization for wireless network aided by intelligent reflecting surface with discrete
	phase shifts,” \textit{IEEE Trans. Commun.,} vol. 68, no. 3, pp. 1838–1851, Mar. 2020.
	
	\bibitem{16}Q. Wu and R. Zhang,~“Intelligent reflecting surface enhanced wireless network via joint active and passive beamforming,” \textit{IEEE Trans. Wireless Commun.},~vol. 18,~no. 11,~pp. 5394-5409,~Nov. 2019.
	
	\bibitem{Tutorial}Q. Wu, S. Zhang, B. Zheng, C. You, and R. Zhang, “Intelligent reflecting surface aided wireless communications: A tutorial,’’ \textit{IEEE Trans. Commun.,} Early Access. 2021
	
	\bibitem{White}N. Rajatheva \textit{et al.}, “White paper on broadband connectivity in 6G,” arXiv preprint arXiv:2004.14247, 2020.
	
	
	\bibitem{8}C. Huang,~A. Zappone,~G. C. Alexandropoulos,~M. Debbah,~and C. Yuen,~“Large intelligent surfaces for energy efficiency in wireless communication,” [Online]. Available: https://arxiv.org/abs/1810.06934.
	
	\bibitem{9}C. Huang,~A. Zappone,~M. Debbah,~and C. Yuen,~“Achievable rate maximization by passive intelligent mirrors,” \textit{in Proc. IEEE ICASSP},~Apr. 2018.
	
	
	
	\bibitem{11}Y. Han,~W. Tang,~S. Jin,~C.-K. Wen,~and X. Ma,~“Large intelligent surface-assisted wireless communication exploiting statistical CSI,” \textit{IEEE Trans. Veh. Technol.},~vol. 68,~no. 8,~pp. 8238–8242,~Aug. 2019.
	
	
	\bibitem{13}G. Zhou,~C. Pan,~H. Ren,~K. Wang,~W. Xu,~and A. Nallanathan,~“Intelligent reflecting surface aided multigroup multicast MISO communication systems,” \textit{IEEE Trans. Signal Process.,} vol. 68, pp. 3236-3251, 2020.
	
	
	\bibitem{Wu11}Q. Wu and R. Zhang,~“Beamforming optimization for wireless network aided by intelligent reflecting surface with discrete phase shifts,’’\textit{ IEEE Trans. Commun.},~vol. 68,~no. 3,~pp. 1838–1851,~Mar. 2020.
	
	\bibitem{Wu22}X. Guan,~Q. Wu,~and R. Zhang,~“Joint power control and passive beamforming in IRS-assisted spectrum sharing,” \textit{IEEE Commun. Lett.},~vol. 24, no. 7, pp. 1553-1557, Jul. 2020.
	
\bibitem{Schoberr}	{X. Yu, D. Xu, D. W. K. Ng, and R. Schober “Power-efficient resource allocation for multiuser MISO systems via intelligent reflecting surfaces,” [Online].~Available: https://arxiv.org/abs/2005.06703.~Accepted at Globecom 2020.}
		
\bibitem{Green_IRS}	{X. Yu, D. Xu, D. W. K. Ng, and R. Schober “IRS-assisted green communication systems: Provable convergence and robust optimization,” 2020, [Online].~Available: https://arxiv.org/abs/2011.06484.}
	
	
	
	%
	
	
	
	\bibitem{18}B. Clerckx,~R. Zhang,~R. Schober,~D. W. K. Ng,~D. I. Kim,~and H. V. Poor,~“Fundamentals of wireless information and power transfer: From RF energy harvester models to signal and system designs,” \textit{IEEE J. Sel. Areas Commun.},~vol. 37,~no. 1,~pp. 4–33,~Jan. 2019.
	
	\bibitem{19} L. Yang,~Y. Zeng,~and R. Zhang,~“Wireless power transfer with hybrid beamforming: How many RF chains do we need?” \textit{IEEE Trans. Wireless Commun.},~vol. 17,~no. 10,~pp. 6972–6984,~Oct. 2018.
	
	\bibitem{19-1}R. Zhang and C. K. Ho,~“MIMO broadcasting for simultaneous wireless information and power transfer,”\textit{ IEEE Trans. Wireless Commun.},~vol. 12,
	no. 5,~pp. 1989–2001,~May. 2013.
	
	\bibitem{21} J. Xu,~L. Liu,~and R. Zhang,~“Multi-user MISO beamforming for simultaneous wireless information and power transfer,” \textit{IEEE Trans. Signal Process.},~vol. 62,~no. 18,~pp. 4798–4810,~Sep. 2014.
	
	\bibitem{222} Q. Shi,~W. Xu,~L. Liu,~and R. Zhang,~“Joint transmit beamforming and receive power splitting for MISO SWIPT systems,” \textit{IEEE Trans. Wireless
		Commun.},~vol. 12,~no. 6,~pp. 3269–3280,~Jan. 2014.
	
	\bibitem{20}Q. Wu and R. Zhang,~“Weighted sum power maximization for intelligent reflecting surface aided SWIPT,” \textit{IEEE Wireless Commun. Lett.,} vol. 9,~no. 5,~pp. 586–590,~May. 2020.
	
	
	
	\bibitem{22}Q. Wu and R. Zhang,~“Joint active and passive beamforming  optimization for intelligent reflecting surface assisted SWIPT under QoS constraints,” \textit{IEEE J. Sel. Areas Commun.}, vol. 38, no. 8, pp. 1735-1748, Aug. 2020.
	
	\bibitem{23}Y. Tang,~G. Ma,~H. Xie,~J. Xu, and X. Han,~“Joint transmit and reflective beamforming design for IRS-assisted multiuser MISO SWIPT systems,”\textit{ ICC 2020 - 2020 IEEE International Conference on Communications (ICC),} Dublin, Ireland, 2020, pp. 1-6.
	
	\bibitem{24}C. Pan,~H. Ren,~K. Wang,~M. Elkashlan,~A. Nallanathan,~J. Wang,~and L. Hanzo,~“Intelligent reflecting surface enhanced MIMO broadcasting for simultaneous wireless information and power transfer,” \textit{IEEE J. Sel. Areas Commun.}, vol. 38, no. 8, pp. 1719-1734, Aug. 2020.
	
		\bibitem{Zargari}S. Zargari, A. Khalili, and R. Zhang, “Energy efficiency maximization via joint active and passive beamforming design for multiuser MISO IRS-aided SWIPT,” \textit{IEEE Wireless Commun. Lett.,} vol. 10, no. 3, pp. 557-561, Mar. 2021. 
	
	\bibitem{Zargari2}	{A. Khalili, S. Zargari, Q. Wu, D. W. K. Ng and R. Zhang “Multi-objective resource allocation for IRS-aided SWIPT,” \textit{IEEE Wireless Commun. Lett.,} 2021.}
			\bibitem{Wu_NOMA} Q. Wu, X. Zhou, and R. Schober, ``IRS-Assisted Wireless Powered NOMA: Is Dynamic Passive Beamforming Really Needed?" arXiv preprint arXiv:2102.08739.
			
	\bibitem{Liu}X. Zhou, R. Zhang,~and C. K. Ho~“Wireless information and power transfer: Architecture design and rate-energy tradeoff,” \textit{IEEE Trans. Commun.}, vol. 61, no. 11, pp. 4754-4767, Nov. 2013.
	
	
	
	\bibitem{Nonlinear}E. Boshkovska,~D. W. K. Ng,~N. Zlatanov,~and R. Schober,~“Practical non-linear energy harvesting model and resource allocation for SWIPT systems,” \textit{IEEE Commun. Lett.},~vol. 19,~no. 12,~pp. 2082-2085,~Dec. 2015.
	
	\bibitem{Schober}D. W. K. Ng,~E. S. Lo,~and R. Schober,~“Wireless information and power transfer: energy efficiency optimization in OFDMA systems,” \textit{IEEE Trans. Wireless Commun.},~vol. 12,~no. 12,~pp. 6352-6370,~Dec. 2013.
	
	\bibitem{Dinkelbach} W. Dinkelbach,~“On Nonlinear Fractional Programming,” \textit{Management Science},~vol. 13,~pp. 492–498,~Mar. 1967.
	
	\bibitem{Bigo}I. P'olik and T. Terlaky,~\textit{Interior Point Methods for Nonlinear Optimization.} Springer,~2010.
	\bibitem{Boyd}S. Boyd and L. Vandenberghe, \textit{Convex Optimization.} Cambridge University
	Press, 2004.
	\bibitem{TWC_Ata}A. Khalili,~M. Robat Mili,~M. Rasti,~S. Parsaeefard,~and D. W. K. Ng,~``Antenna selection strategy for energy efficiency maximization in uplink OFDMA networks: A multi-objective approach,” \textit{IEEE Trans Wireless Commun.},~vol. 19,~no. 1,~pp. 595-609,~Jan. 2020.
	
	\bibitem{Jorswieck}A. Zappone,~E. Björnson,~L. Sanguinetti,~and E. Jorswieck,~“Globally optimal energy-efficient power control and receiver design in wireless networks,” \textit{IEEE Trans. Signal Process.},~vol. 65,~no. 11,~pp. 2844-2859,~Jan. 2017.
	
	
	\bibitem{B5G}G. Yu, X. Chen, C. Zhong, D. W. K. Ng, and Z. Zhang, “Design, analysis and optimization of a large intelligent reflecting surface aided B5G cellular Internet of Ihings,” \textit{IEEE Internet of Things Journal}, vol. 7, no. 9, pp. 8902-8916, Sept. 2020.
	
	
	
	
	\bibitem{CVX}M. Grant and S. Boyd. CVX: MATLAB software for disciplined convex programming. [Online]. Available: http:cvxr.com/cvx.
	
	
	
	
	\bibitem{32}Y. Sun,~P. Babu,~and D. P. Palomar,~“Majorization-Minimization algorithms in signal processing,~communications,~and machine learning,”
	\textit{IEEE Trans. Signal Process.},~vol. 65,~no. 3,~pp. 794-816,~Feb. 2017.
	
	\bibitem{33}D. W. K. Ng,~E. S. Lo, and R. Schober,~“Robust beamforming for secure communication in systems with wireless information and power transfer,” \emph{IEEE Trans. Wireless Commun.},~vol. 13,~no. 8,~pp. 4599-4615,~Aug. 2014.
	
	\bibitem{penalty}D. Xu,~X. Yu,~Y. Sun,~D.W.K. Ng,~and R. Schober,~“Resource allocation for IRS-assisted full-duplex cognitive radio systems,” \textit{IEEE Trans. Commun.}, [Online]. Available: https://arxiv.org/abs/2003.07467,~Mar. 2020.
	
	\bibitem{bjorson}	{A.~Zappone,~E.~Björnson,~L.~Sanguinetti,~and E.~Jorswieck,~“Globally optimal energy-efficient power control and receiver design in wireless networks,” \textit{IEEE Trans. on Signal Process.}, vol. 65, no. 11, pp. 2844-2859, Jun. 2017.}
	
		\bibitem{amplitud}	{M. M. Zhao and Q. Wu, M. J. Zhao, and R. Zhang,~“Exploiting amplitude control in intelligent reflecting surface aided wireless communication with imperfect CSI,” \textit{IEEE Trans. Commun}, [Online]. Available: https://arxiv.org/abs/2005.07002, 2020.}

	
		\bibitem{Guo}	{H. Guo, Y.-C. Liang, J. Chen, and E. G. Larsson, “Weighted sum-rate optimization for intelligent reflecting surface enhanced wireless networks,” [Online]. Available: https://arxiv.org/abs/1905.07920, 2020.}


		\bibitem{Marks} 	{B. R. Marks and G. P. Wright, “A general inner approximation algorithm for nonconvex mathematical programs,” \textit{Operations Research,} vol. 26, no. 4, pp. 681–683, Jul. 1978.}


		\bibitem{Razaviyayn}	{ M. Razaviyayn, M. Hong, and Z.-Q. Luo, “A unified convergence analysis of block successive minimization methods for nonsmooth optimization,” SIAM J. Optim., vol. 23, no. 2, pp. 1126–1153, 2013.}
		
		
		
		\bibitem{TangTang} { W. Tang, et al., “Wireless communications with reconfigurable intelligent surface: Path loss modeling and experimental measurement,” arXiv preprint arXiv:1911.05326, 2019.~Accepted by \textit{IEEE Trans. Wireless Commun}.}
		
		\bibitem{WanWan} { T. J. Cui, M. Q. Qi, X. Wan, J. Zhao, and Q. Cheng, “Coding metamaterials, digital metamaterials and programmable metamaterials,”\textit{Light: Science \& Applications,} vol. 3, no. 10, Oct. 2014.}
		
		\bibitem{LeroseyLerosey}{ N. Kaina, M. Dupré, G. Lerosey, and M. Fink, “Shaping complex microwave fields in reverberating media with binary tunable metasurfaces,” \textit{Sci. Rep.,} vol. 4, p. 6693, Oct. 2014.}
		
		\bibitem{Space}{L. Zhang et al., “Space-time-coding digital metasurfaces,”\textit{ Nat. Commun.,} vol. 9, no. 1, p. 4334, Oct. 2018.}
		\bibitem{Ata_WCNC}A. Khalili and D. W. K. Ng, “Energy and spectral efficiency tradeoff in OFDMA networks via antenna selection strategy,” \textit{in Proc. IEEE WCNC}, Seoul, Korea (South), 2020, pp. 1-6.
		
		\bibitem{Ata_ICC}A. Khalili, M. R. Mili, and D. W. K. Ng, “Performance trade-off between uplink and downlink in full-duplex communications,” \textit{in Proc. IEEE ICC}, Dublin, Ireland, 2020, pp. 1-6.
		
\end{thebibliography}
\end{document}